\documentclass[12pt]{article}
\usepackage{caption}
\usepackage{graphicx} 
\usepackage{amssymb}
 \title{Neutrino production from photo-hadronic interactions of the 
 gamma flux from Active Galactic Nuclei with their gas content} 
 \author{J.C. Arteaga-Vel\'azquez}
 \date{}

\begin{document}

   \maketitle
   \vspace{-3pc}
   \begin{center}
   \textit{Instituto de F\'\i sica y Matem\'aticas, Universidad Michoacana,
    Edificio C3, Cd. Universitaria, 58040 Morelia, Michoacan, Mexico}\\
    Email: arteaga@ifm.umich.mx
   \end{center}

   \begin{abstract}
   The diffuse neutrino flux from FRI and BL Lac type 
   galaxies generated from interactions of their own $\gamma$ radiation with the
   gas and dust at the sources is reported. This $\nu$-production channel has not been 
   studied in detail up to now. The calculations are based on individual 
   estimations of the neutrino flux in two nearby AGN's: Centaurus A and M87,
   assuming the validity of the AGN unification model. The predictions for
   Centaurus A and M87 involved the parameterization of the measured gamma-ray 
   luminosities and the modeling of the material of the galaxies both 
   based on observations performed by several detectors. 
   No hadronic origin for the TeV photons is assumed. The results show 
   that, although the corresponding $\nu$ flux ($E^2 \Phi_{\nu + \bar{\nu}} 
   \lesssim 10^{-13} \, \mbox{s}^{-1} \, \mbox{sr}^{-1} \, \mbox{GeV} \, \mbox{cm}^{-2}$) 
   is not competitive at high-energies ($\epsilon_\nu \gtrsim 1 \, \mbox{TeV} $) 
   with that from hadronic models, FRI and BL Lac galaxies with $\gamma$ emission
   should be already contributing to the diffuse flux of neutrinos in the universe.    
   \end{abstract}

   \vspace{1pc}

   \vspace{1pc}
   \textit{This is an author-created, un-copyedited version of an article published in 
   Astroparticle Physics. Astroparticle Physics is not responsible for any errors 
   or omissions in this version of the manuscript or any version derived from it. The 
   definitive published authenticated version is available at doi:
   10.1016/j.astropartphys.2012.07.002}

  \section{Introduction}

   According to particle physics, high-energy neutrino production should be a common phenomenon
  in several places in the universe. At high-energies an important mechanism of 
  neutrino production is by means of pion photoproduction reactions induced by interactions of
  cosmic rays with the matter and radiation in the universe, the so called hadronic model. 
  If this picture turns out to be true,  a deep connexion among the gamma ray, neutrinos and 
  cosmic rays could be established, which can serve to explore the same object in different 
  windows and learn about the acceleration mechanism of the most energetic particles in the 
  universe. 
  
  By using hadronic interaction models,  predictions of high-energy neutrino
  fluxes for several astrophysical objects can be estimated \cite{Becker}, but they rely on
  the strong assumption that cosmic rays are accelerated at the sources, which
  still has to be proven. However, independently of the above scenario and the
  presence of cosmic rays,  there should be secondary mechanisms of neutrino
  production which may be already at work at astrophysical environments that
  emit $\gamma$ radiation. Among such mechanisms are muon pair creation
  \cite{Razzaque06} and charged pion production \cite{Menon09} by $\gamma$-ray
  collisions with the ambient radiation, thermal photoproduction of neutrino
  pairs ($\gamma e^{\pm} \rightarrow e^{\pm} + \nu + \bar{\nu}$)
  \cite{Beaudet66} and photointeractions of the $\gamma$-rays of the source
  with stellar atmospheres and with the gas and dust surrounding the emission 
  region. Although the associated probabilities for these processes are
  lower than that for the hadronic channel, these $\gamma$-ray mediated 
  collisions should be contributing to the background of high-energy neutrinos
  in the universe. 

  In consequence, several astrophysical gamma ray sources should be producing
  continuously neutrinos through the referred secondary  reactions, for
  example, Active Galactic Nuclei. With the arrival of the \textit{Fermi}-LAT
  telescope, it has been common to find gamma ray emissions from these kind
  of sources \cite{Fermi-list, Abdo09}, for example, FRI \cite{FRI} and BL
  Lacs \cite{Bllac}, which are some special cases of Radio Loud Galaxies
  \cite{Urry}. Then it is natural to assume a diffuse flux of high-energy
  neutrinos in the universe produced by the gamma interactions of these
  populations of AGN's with their own ambient radiation and material. In this
  paper, the diffuse $\nu$ flux from FRI and BL Lac objects produced in
  $\gamma P$ interactions of high-energy photons with the gas and dust at
  source is calculated and compared with some AGN 
  predictions derived from hadronic models of cosmic rays. This channel has
  not been studied in detail up to now.
  
  Only photo-hadronic interactions will be considered for the
  present work. The contribution from direct $\mu$-pair production with 
  subsequent muon decay will be neglected, since its cross section 
  is small in comparison with that for the hadronic case. In the former, 
  $\sigma \sim \mu b$ \cite{Halzen86} and, in the latter, of the order of 
  $mb$ \cite{PDG10}.

  In section 2, the expression for the estimation of the diffuse neutrino flux 
  as a function of the photon luminosity of the sources is presented. Along this paper,
  it is assumed a standard cosmological scenario with $h = 0.72$, $\Omega_\Lambda = 0.74$ 
  and $\Omega_m = 0.26$. For the calculations, Centaurus A (CenA) and M87  are taken as 
  representative objects. Both AGN's are classified as FRI type galaxies, which in the 
  standard AGN unification scenario \cite{Urry}, correspond to misaligned 
  BL Lac sources. The estimated photon luminosities for a single source are shown in 
  section 3. They are obtained from individual fits to the photon luminosities of
  Centaurus A and M87 measured with telescopes at Earth and on space.

  A relevant quantity for the estimation of the diffuse neutrino flux is the 
  gamma-ray luminosity function (GLF) of the sources. The GLF's for FRI and BL Lac 
  galaxies are presented in section 4. The functions are normalized according to the
  results of the latest \textit{Fermi}-LAT surveys \cite{Fermi-list, Abdo10} as it 
  will be described.

  The models of matter distribution for FRI and BL Lac galaxies are based on
  observations of Centaurus A and M87. They are introduced in section 5.
  In section 6, the neutrino yield for photoproduction reactions is shown. This
  quantity is calculated using the Monte Carlo program SOPHIA v2.01 \cite{Sophia}.
  The formula to calculate the total $\gamma P$ hadronic cross section is also
  presented in this section. The expression is taken directly from the results
  of the fits of the COMPETE Collaboration \cite{PDG10, Compete02}.

  The estimated diffuse neutrino fluxes are finally shown in section 7. The 
  results are discussed and compared with both limits set by modern 
  neutrino observatories and AGN estimations based on hadronic models. 
  Uncertainties in the flux associated with the observed distribution of 
  photon indexes for BL Lac and FRI type objects will be considered in 
  this part of the paper. Conclusions are presented in section 8.

 \section{The $\gamma$-ray induced neutrino flux}

 \subsection{Production at source}

  Consider an AGN with a $\gamma$-ray emission characterized by a photon 
  spectral luminosity $L_\gamma (\epsilon_\gamma)$, where 
  $\epsilon_\gamma$ is the photon energy, both measured at the 
  source frame. Let's assume, for simplicity, that the position of the 
  gamma-ray source is located at  the core of the AGN and that both the 
  gas and dust of the host galaxy are composed of protons, which are at 
  rest. Then, the neutrino spectral luminosity,
  \begin{equation}
      L_\nu(\epsilon_\nu) = 
      \frac{dN_\nu}{dt d\epsilon_\nu},
      \label{eq1}
  \end{equation} 
  produced by the interaction of the gamma-photons with the intergalactic 
  material of the AGN along their path can be estimated as
  \begin{equation}
    L_\nu(\epsilon_\nu) d\epsilon_\nu = 
     \bar{\Sigma}_{\small H} \int_{\epsilon_{\gamma, i}}^{\epsilon_{\gamma, f}}
     \sigma_{\gamma P}(\epsilon_\gamma) 
    Y^{\gamma P \rightarrow \nu}(\epsilon_\gamma, \epsilon_\nu)
    L_\gamma (\epsilon_\gamma) d\epsilon_\gamma.
    \label{eq2}
  \end{equation}  
  Here, $\bar{\Sigma}_{\small H}$ is the angle-averaged column density for target 
  protons (in units of $\mbox{cm}^{-2}$), $\sigma_{\gamma P}$ (measured in $\mbox{cm}^{2}$)
  is the $\gamma P$ cross section at a photon energy $\epsilon_\gamma$  and 
  $Y^{\gamma P \rightarrow \nu}(\epsilon_\gamma, \epsilon_\nu)$ is the neutrino yield,
  which is defined as the number of neutrinos with energy in the region $d\epsilon_\nu$ 
  around $\epsilon_\nu$  that a photon with energy in the interval $[\epsilon_\gamma, 
  \epsilon_\gamma + d\epsilon_\gamma]$ produces after its collision with a
  proton at rest. In the above expression $\epsilon_\gamma$ is in
  $\mbox{TeV}$, while particle spectral luminosities are in units
  $\mbox{s}^{-1} \cdot \mbox{TeV}^{-1}$.

  In equation (\ref{eq2}), the limits of integration are set in the high-energy 
  regime: $\epsilon_{\gamma, i} = 10^{-0.8}  \, \mbox{GeV}$ and $\epsilon_{\gamma, f} = 10^{6} \, 
  \mbox{GeV}$, while the neutrino energy covers the interval from  $\epsilon_{\nu, i} = 10^{-5}  \, 
  \mbox{GeV}$ to $\epsilon_{\nu, f} = 10^{6} \, \mbox{GeV}$. Note that, in
  $\gamma P$ interactions, the gamma energy threshold for pion photoproduction
  is $\epsilon_{\gamma, th} = m_\pi(m_\pi c^2 + 2m_p c^2)/2m_p \approx
  10^{-0.8} \, \mbox{GeV}$, at the laboratory system \cite{Propagation-uhecr}.  

   Several assumptions have been made to derive equation (\ref{eq2}). First, to 
  give a rough estimation of $L_\nu(\epsilon_\nu)$ at source, the photon spectral 
  luminosity was considered to be isotropic and constant inside the AGN. It was set 
  equal to the measured value at source as derived  from Earth observations. In fact, 
  this quantity is associated with the $\gamma$ photons which escape without 
  interactions from the AGN. In any case, since corrections due
  to absorption effects should increase the photon spectral luminosity and
  neutrinos can emerge from regions where photons may not, the quantity
  evaluated in (\ref{eq2}) should be taken as very conservative. For a more
  realistic estimation, both the position and view-angle dependence of the AGN
  luminosity should be considered.

  In the above equation, the influence of the intergalactic magnetic field, 
  the photon background and the gas at the source on the absorption  and
  energy loss of the parent particles in the $\nu$-production chain has been 
  neglected. These phenomena may play an important role close to the AGN 
  core (see, for example, \cite{Rachen}) reducing the neutrino luminosity. 
  In the worst scenario, the high-energy neutrino production could be suppressed. 
  To estimate in this extreme case the neutrino luminosity, equation (\ref{eq2}) 
  will be alternatively evaluated applying a cut around the center of the
  corresponding gamma-ray source. 

  The assumption that the target protons are at rest could work for the 
  hot, warm and cold gasses of the interstellar medium (ISM) in the AGN, 
  but not for the gas moving outwards at high relativistic speeds along 
  the jets. For example, the temperature of the gas supply for the central 
  engine of some AGN's has been estimated to be of the order of $\mbox{keV}$ 
  \cite{Allen06, Evans06, centA-Evans04}, in such a medium, the mean kinetic
  energy expected for protons is roughly of the same order of magnitude, 
  which results to be small in comparison with the gamma energies considered
  in this work ($\epsilon_{\gamma} > 10^{-0.8} \, \mbox{GeV}$). In the case of
  the jets, material with bulk Lorentz factors $\Gamma \sim \mathcal{O}(10)$
  seems to be present in several AGN's \cite{Kellerman04, Lister09}, which
  could imply proton kinetic  energies of the order of $\Gamma m_p c^2 \sim 10
  \, \mbox{GeV}$ (if ultra-high-energy acceleration is not considered
  \cite{Aharonian02}), that are not negligible in comparison  with
  $\epsilon_{\gamma}$. This situation is important for BL Lac galaxies, where
  the jets are closely aligned with the line of sight. Here photointeractions 
  with the relativistic gas flowing along the jets are expected to have a
  major contribution  in equation (\ref{eq2}). 

   If both the gamma flux and the relativistic medium are moving outwards, it 
  results in a smaller amount of energy available for neutrino production and, 
  in consequence, in a  lower $\nu$ luminosity than for the situation with 
  stationary gas. The estimation of the neutrino spectral luminosity in this
  case is carried out by considering the interaction of gamma-photons with a
  gas characterized by a Lorentz factor $\Gamma = 10$, both moving in the same
  direction. Calculations of the neutrino luminosity are first performed at
  the rest frame of the gas using formula (\ref{eq2}). The result is then
  Lorentz transformed to the laboratory reference system through the
  expression
  \begin{equation}
      L_\nu \left[ \epsilon_\nu = 
      \frac{\epsilon^{\prime}_\nu} {\Gamma (1 - \beta)} \right] = 
      (1 - \beta) L^{\prime}_\nu(\epsilon^{\prime}_\nu),
      \label{eq3}
  \end{equation}
  with $\beta = \sqrt{1 - 1/\Gamma^{2}}$.  The prime stands for quantities 
  measured at the gas rest frame. A similar equation is employed to evaluate
  the photon spectral luminosity at the reference frame of the jet. 

  It is clear that more realistic estimations should also include detailed
  modeling of the photon's and proton's angular distributions of their
  momenta. But for the purpose of giving a rough estimation of the neutrino 
  flux from AGN's these distributions will be left out of the present 
  calculations.

  \subsection{Diffuse neutrino flux}

  The extragalactic neutrino background observed at Earth due to FRI objects
  and BL Lac galaxies was calculated by means of the relation 
  \begin{equation}
      \frac{d\Phi_\nu(\epsilon^{\circ}_\nu)}{d\Omega^{\circ}} = 
      \frac{c}{4\pi} \int_{0}^{z_{max}} dz \frac{1}{H(z)}
      \int_{\log_{10}\mathcal{L}_{\gamma}^{min}}^{\log_{10}\mathcal{L}_{\gamma}^{max}}
      L_\nu[\mathcal{L}_\gamma, \epsilon^{\circ}_\nu(1+z)] \cdot
      \rho_\gamma(\mathcal{L}_\gamma, z)\cdot
      d(\log_{10}\mathcal{L}_\gamma)
      \label{eq4}
  \end{equation}
  where $ L_\nu[\mathcal{L}_\gamma, \epsilon^{\circ}_\nu(1+z)]$ is the neutrino
  spectral luminosity of an individual AGN with integrated gamma-ray 
  luminosity 
  \begin{equation}
     \mathcal{L}_\gamma = \mathcal{L}_\gamma(\epsilon_{\gamma_1},
     \epsilon_{\gamma_2}) 
     = \int_{\epsilon_{\gamma_1}}^{\epsilon_{\gamma_2}}
     \epsilon_\gamma \cdot L_\gamma (\epsilon_\gamma) d\epsilon_\gamma,
     \label{eq5}
  \end{equation}  
  and located at a redshift $z$, with $\epsilon_\nu =
  \epsilon^{\circ}_\nu(1+z)$ representing the neutrino energy at source and
  $\epsilon^{\circ}_\nu$ the corresponding particle energy at Earth after 
  adiabatic energy losses due to redshift. 
  $\rho_\gamma(\mathcal{L}_\gamma, z)$ is the gamma-ray luminosity function (GLF) 
  of the AGN sources per comoving volume $dV_c$ and interval 
  $d(\log_{10}\mathcal{L}_\gamma)$, in units of $\mbox{cm}^{-3}$, while
  \begin{equation}
     c/H(z) = (c/H_0) [1 - \Omega_m + \Omega_m (1 + z)^3]^{-1/2},
     \label{eq6}
  \end{equation}  
  where $H(z)$ is the Hubble parameter at redshift $z$ and $c/H_0 = 1.28
  \times 10^{28} \, \mbox{cm}$ \cite{PDG10} is the Hubble length. A flat 
  cosmological model dominated by vacuum energy with $\Omega_\Lambda = 
  1 - \Omega_m = 0.74$ is assumed. The neutrino background flux
  is given in units $\mbox{s}^{-1} \cdot \mbox{sr}^{-1} \cdot \mbox{GeV}^{-1}
  \cdot \mbox{cm}^{-2}$.

 \section{The gamma ray luminosity for a single source}

 To evaluate $L_\gamma (\epsilon_\gamma)$ for FRI radio galaxies, the
 gamma-ray fluxes from Centaurus A and M87 are employed. These sources will be
 considered as representative ones. Of course, for a more realistic
 calculation the measured distributions of the spectral indices and integrated
 luminosities for this kind of sources should be included \cite{Fermi-Lat-spectra}.
 
 Centaurus A and M87 are nearby FRI type objects which have been detected in
 the MeV-TeV energy regime. The source of the gamma-ray emission has not been
 identified yet. Astronomical observations indicate that the $\gamma$
 radiation could be produced in the nucleus, the inner jet or radio lobes of
 the sources \cite{centA-hess09, centA-fermi10, m87-fermi09}. More stringent
 limits come from combined multi-wavelength observations of M87, which point
 out that the gamma-ray production site could be located at the nucleus of the 
 AGN \cite{m87-multi09}. The mechanism behind this emission is also
 unknown. Both leptonic and hadronic models have been invoked to explain the
 $\gamma$-ray production, however, modern observations are still not able to
 exclude among the different scenarios (see, for example, \cite{centA-fermi10,
 m87-multi09}).

 The closest FRI radio galaxy to the Earth is Centaurus A. It is located at an
 approximate distance of $3.4 \, \mbox{Mpc}$ \cite{Quillen06}. Measurements of
 the $\gamma$-ray flux in the interval $0.1 - 1 \, \mbox{GeV}$ have been
 performed with the EGRET detector on board of the Compton  Gamma-Ray
 Observatory \cite{centA-egret98, centA-egret99}. In the $0.1 - 30  \,
 \mbox{GeV}$ energy range, data has been collected with the \textit{Fermi}-LAT
 instrument \cite{centA-fermi10} and in the very high-energy region
 ($\epsilon^{\circ}_\gamma >100 \, \mbox{GeV}$), with the telescopes of the
 H.E.S.S. experiment \cite{centA-hess09}. Centaurus A has an integrated
 $\gamma$-ray flux of $\Phi_\gamma(\epsilon^{\circ}_\gamma >100 \, \mbox{MeV})
 = (1.50 \pm 0.45) \times 10^{-7} \, \mbox{ph} \cdot \mbox{cm}^{-2} \cdot
 \mbox{s}^{-1}$ \cite{centA-fermi10}.

 On the other hand, M87 is found at a distance of $16.7 \, \mbox{Mpc}$ from
 the Earth \cite{m87-multi09}. Its gamma-ray flux has been measured with the
 \textit{Fermi}-LAT telescope in the range of $0.2$ to $32 \, \mbox{GeV}$ and
 with HEGRA \cite{m87-hegra}, HESS \cite{m87-hess} and  VERITAS
 \cite{m87-veritas08} in the energy regime of $0.1 - 30 \, \mbox{TeV}$. Above
 $100 \, \mbox{MeV}$, the integrated flux of M87 is $(2.45 \pm 0.63) \times
 10^{-8} \, \mbox{ph} \cdot \mbox{cm}^{-2} \cdot \mbox{s}^{-1}$, almost an
 order of magnitude smaller than that for Centaurus A \cite{m87-fermi09}. 

 Now, let's be $\Phi_\gamma(\epsilon^{\circ}_\gamma) =
 dN_\gamma/dt^{\circ}d\epsilon^{\circ}_\gamma dA^{\circ}$ the differential
 photon spectrum (in units of $\mbox{s}^{-1} \cdot \mbox{TeV}^{-1} \cdot
 \mbox{cm}^{-2}$) detected at Earth from Centaurus A or M87. Suppose a steady
 and isotropic emission, the photon spectral luminosity of the AGN, corrected
 for redshift energy losses and attenuation due to interactions with the
 background radiation, is given by 
 \begin{equation}
   L_\gamma (\epsilon_\gamma) = 
   4\pi D^2_L \cdot
   \frac{\Phi_\gamma(\epsilon_\gamma)}{(1+z)^2} \cdot 
   \mbox{e}^{\tau(\epsilon_\gamma, z)},
   \label{eq7}
 \end{equation}
 where $z$ and $D_L$ (in $\mbox{cm}$) are the redshift and luminosity
 distance to the source, both related through
 \begin{equation}
   D_L(z)  = (1 + z) \int_{0}^{z} c\,dz/H(z),
   \label{eq8}
  \end{equation}  
 and $\tau(\epsilon_\gamma, z)$ is the $\gamma \gamma$ optical depth for
 a photon traveling from the source with initial energy $\epsilon_\gamma$.
 For this work, $\tau(\epsilon_\gamma, z)$ is extracted from
 \cite{gg-optical-depth}.

 \begin{figure}[!t]
 \centering
 \includegraphics[width=3.2in]{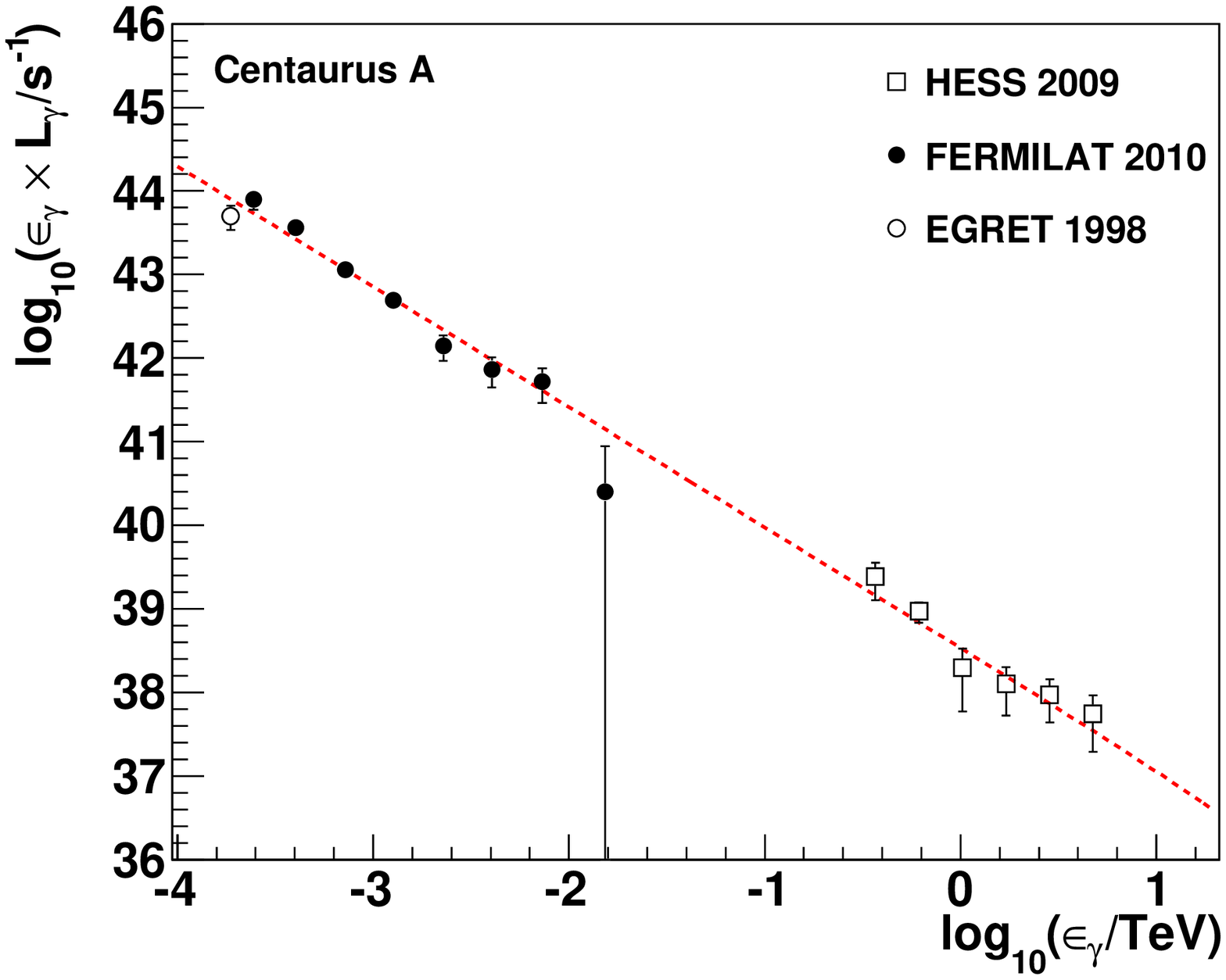}
 \includegraphics[width=3.2in]{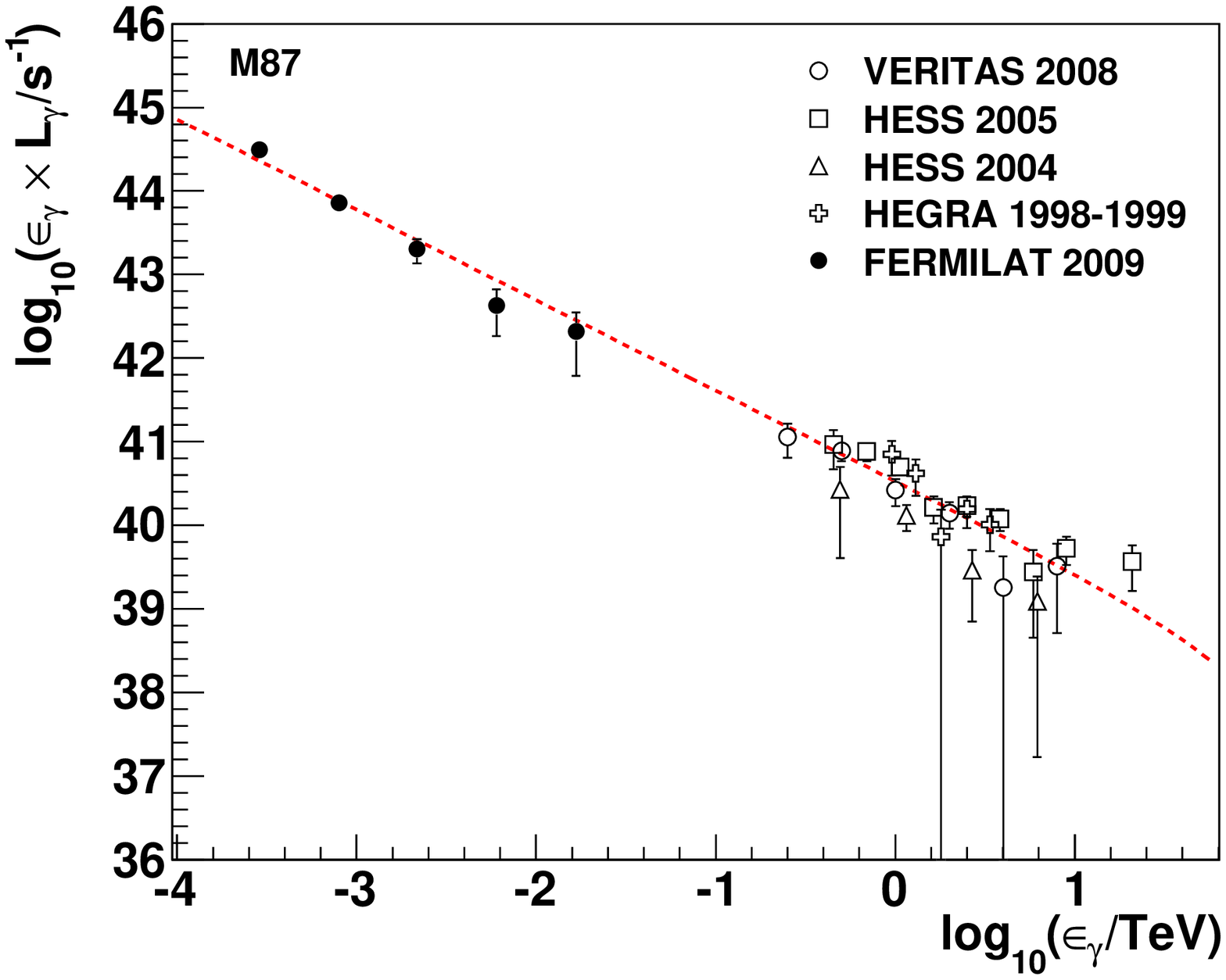}
  \caption{Photon spectral luminosities for Centaurus A (left) and M87 (right)
 derived from $\gamma$-ray measurements performed with different experiments
 (data points, see text). The results of the fits with formula (\ref{eq9}) are
 shown in each case (segmented lines).}
 \label{fig1}
\end{figure}

 The photon spectral luminosity for Centaurus A, multiplied by
 $\epsilon_\gamma$, is presented in Fig. \ref{fig1}. It is calculated using
 $D_L = 3.4 \, \mbox{Mpc}$ \cite{Quillen06} in equation (\ref{eq7}) along with
 the differential photon spectrum as derived from measurements with EGRET
 \cite{centA-egret98}, HESS \cite{centA-hess09}  and \textit{Fermi}-LAT
 \cite{centA-fermi10}. In the same figure, the result for M87 is shown. Here
 $D_L = 16.7 \, \mbox{Mpc}$ \cite{m87-multi09} and the data from HEGRA
 \cite{m87-hegra}, HESS \cite{m87-hess}, VERITAS \cite{m87-veritas08} and
 \textit{Fermi}-LAT \cite{m87-fermi09} are taken into account.

 To describe the photon spectral luminosities, each graph in Fig. \ref{fig1}
 is fitted using a  power-law function with a cut-off at high-energies,
 \begin{equation}
   L^{\mbox{\tiny{fit}}}_\gamma (\epsilon_\gamma) = 
   b \cdot 
   \left[\frac{\epsilon_\gamma}{\mbox{TeV}}\right]^{a} \cdot
    e^{-(\epsilon_\gamma/10^{2} \, \mbox{\footnotesize{TeV}})},
   \label{eq9}
 \end{equation}
 where $a$ and $b$ stand for the fit parameters. The cut is introduced
 considering the possibility that the gamma-ray emission could not be of
 hadronic but leptonic origin (see, for example, \cite{Reynoso10}). The
 results of the fit are shown in table \ref{tab1}.

 \begin{table}[!t]
 \begin{center}
 \caption{Values of the fit parameters $a$ and $b$ of formula (\ref{eq9})
 for Centaurus A and M87.}
 \begin{tabular}{l|cc}
 \hline
 AGN  &$\log_{10} [b/\mbox{s}^{-1} \cdot \mbox{TeV}^{-1}]$ & $a$\\
 \hline
 Centaurus A  & $38.53 \pm 0.09$ & $-2.44  \pm 0.03$ \\                
 M87          & $40.53 \pm 0.03$ & $-2.08  \pm 0.02$\\
 \hline
 \end{tabular}
 \label{tab1}
 \end{center}
 \end{table}

 The integrated luminosity above $100 \, \mbox{MeV}$ for Centaurus A and M87 
 is evaluated substituting the values of the fit parameters and formula
 (\ref{eq9}) in equation (\ref{eq5}). In the former, it is obtained
 $\mathcal{L}_\gamma (\epsilon_\gamma >100 \, \mbox{MeV}) = 7.2 \times 10^{40}
 \, \mbox{erg} \cdot \mbox{s}^{-1}$, which is more than an order of magnitude
 lower than in the case of M87 for which $\mathcal{L}_\gamma(\epsilon_\gamma
 >100 \, \mbox{MeV}) = 1.1 \times 10^{42} \, \mbox{erg} \cdot \mbox{s}^{-1}$.

 Using the photon spectral luminosities of Centaurus A and M87 as a reference,
 the corresponding luminosity for a single source characterized by an integral
 luminosity $\mathcal{L}_\gamma$ can be calculated by means of the formula
 \begin{equation}
   L_\gamma (\epsilon_\gamma) = L^{\mbox{\tiny{fit}}}_\gamma (\epsilon_\gamma) \cdot
   \left[ \frac{\mathcal{L}_\gamma}{\mathcal{L}^{\mbox{\tiny{fit}}}_\gamma} \right],
   \label{eq10}
 \end{equation} 
 where $\mathcal{L}^{\mbox{\tiny{fit}}}_\gamma$ is evaluated using (\ref{eq9})
 in equation (\ref{eq5}).

 The luminosities here derived from Centaurus A and M87 will be also applied 
 for BL Lac galaxies, which is justified by appealing to the AGN unification
 scheme \cite{Urry}. In this case, it will be assumed that the gamma ray
 emission from FRI and BL Lac objects has a common origin and that the
 respective $\gamma$-ray fluxes are isotropic and of equal magnitude.

 \section{The gamma-ray luminosity functions}

  \subsection{The GLF for FRI galaxies}
 
  The gamma-ray luminosity function (GLF) for FRI radio galaxies used in this work
  is taken from reference \cite{GLF-inoue11}. In that paper, the GLF function
  is obtained exploiting the correlation between radio and gamma-ray
  luminosities exhibited by FRI and FRII radio galaxies with $\gamma$-ray
  emission. Once established  such a correlation, the radio luminosity
  function (RLF) determined in  \cite{RLF-willott01} for  radio loud galaxies
  is converted to the GLF, which is later normalized according to the source
  count distribution of  $\gamma$-ray loud radio  galaxies measured with
  \textit{Fermi}-LAT \cite{Abdo10}.

 \begin{table}[!t]
 \begin{center}
 \caption{Values of the parameters of the RLF \cite{GLF-inoue11, RLF-willott01}.}
 \begin{tabular}{ccccccccc}
 \hline
 $\log_{10}(\rho_I/\mbox{cm}^{-3})$ && $\alpha_I$ &&
 $\log_{10}(\mathcal{L}_I/\mbox{W} \cdot \mbox{Hz}^{-1} \cdot \mbox{sr}^{-1})$
 && $z_I$ && $k_I$ \\
 \hline
 $-80.991$ && $0.586$ && $26.48$ && $0.710$ && $3.48$\\
 \hline
 \end{tabular}
 \label{tab2}
 \end{center}
 \end{table}

  The GFL is given by the following expression \cite{GLF-inoue11}:
  \begin{equation}
    \rho_\gamma(\mathcal{L}_\gamma, z) = \kappa 
    [ d(\log_{10}\mathcal{L}_{{}_{151 \, \mbox{\tiny MHz}}})/d(\log_{10}\mathcal{L}_\gamma)]
    \rho_r(\mathcal{L}_{{}_{151 \, \mbox{\tiny MHz}}}, z).
   \label{eq11}
  \end{equation}
  where $\kappa = 0.081 \pm 0.011$ is the normalization factor,
  $\mathcal{L}_\gamma$ is the integrated  $\gamma$-ray  luminosity in the
  interval $10^{-4} - 10^{-2} \, \mbox{TeV}$ and  $\mathcal{L}_{{}_{151 \,
  \mbox{\tiny MHz}}}$ is the  151 MHz monochromatic luminosity in units of
  $\mbox{W}/\mbox{Hz} \cdot \mbox{sr}$  \cite{RLF-willott01}. On the other
  hand, $\rho_r(\mathcal{L}_{{}_{151 \, \mbox{\tiny MHz}}}, z)$ is the 151 MHz
  RLF for FRI galaxies, measured in $\mbox{cm}^{-3}$ per interval of comoving
  volume and $\log_{10}\mathcal{L}_{{}_{151 \, \mbox{\tiny MHz}}}$. This
  quantity is determined in reference \cite{RLF-willott01} and it is shown
  below:
  \begin{eqnarray}
    \rho_r(\mathcal{L}_{{}_{151 \, \mbox{\tiny MHz}}}, z) =
    \left\{ \begin{array}{l}
           \eta(z) \rho_I (\frac{\mathcal{L}_{{}_{151 \, \mbox{\tiny MHz}}}}{\mathcal{L}_I})^{-\alpha_I}
           \mbox{exp}(\mathcal{L}_{{}_{151 \, \mbox{\tiny MHz}}}/\mathcal{L}_I)
           (1+z)^{k_I}  \, \, \, \, \, z \leq z_I,\\
           \eta(z) \rho_I (\frac{\mathcal{L}_{{}_{151 \, \mbox{\tiny MHz}}}}{\mathcal{L}_I})^{-\alpha_I}
           \mbox{exp}(\mathcal{L}_{{}_{151 \, \mbox{\tiny MHz}}}/\mathcal{L}_I)
           (1+z_I)^{k_I}  \, \, \, \, z > z_I. \\
           \end{array}
     \right.
    \label{eq12}
  \end{eqnarray}
  The parameters of the RLF are presented in table \ref{tab2}. In the above 
  equation  \cite{GLF-inoue11}
  \begin{equation}
   \eta(z) = \frac{d^2V_{{}_W}/d\Omega dz}{d^2V_c/d\Omega dz}
   \label{eq13}
  \end{equation}
  is a cosmological conversion factor to compensate for the fact that in 
  \cite{RLF-willott01} a flat universe with $\Omega_m = 1$ and $\Omega_\Lambda
  = 0$ is employed. In formula (\ref{eq13})
  \begin{equation}
   \frac{d^2V_{{}_W}}{d\Omega dz} = \frac{c^3 z^2 (2+z)^2}{4H_{0,W}^3(1+z)^3},
   \label{eq14}
  \end{equation}
  with $c/H_{0,W} = 1.85 \times 10^{28} \, \mbox{cm}$ and
  \begin{equation}
   \frac{d^2V_c}{d\Omega dz} = \frac{c [D_L(z)]^2}{H(z)(1+z)^2}.
   \label{eq15}
  \end{equation}
  The relation between $\mathcal{L}_{{}_{151 \, \mbox{\tiny MHz}}}$ and 
  $\mathcal{L}_\gamma$ is established in \cite{GLF-inoue11} through their
  correlation with the 5 GHz radio luminosity, $\mathcal{L}_{{}_{5 \, 
  \mbox{\tiny GHz}}}$, where $\mathcal{L}_f = f \cdot \epsilon_f dN/dt df$.
  Here, $f$ is the frequency of the radiation and $\epsilon_f$, the energy of
  the associated photon. First, the 151 MHz monochromatic luminosity is
  transformed to $\mathcal{L}_{{}_{5 \, \mbox{\tiny GHz}}}$
  assuming a spectral index  $\alpha_r = 0.8$ for all radio galaxies 
  \cite{RLF-willott01}, i.e. $S_{r}(f) \propto f^{-\alpha_r}$, where $S_r$ is
  the radio flux density at frequency $f$, and then the radio luminosity at $5 \,
  \mbox{GHz}$ is converted to the integrated gamma-ray luminosity using the
  expression \cite{GLF-inoue11}
   \begin{equation}
    \log_{10}(\mathcal{L}_\gamma/\mbox{erg} \cdot \mbox{s}^{-1})
    = (-3.90 \pm 0.61) + (1.16 \pm 0.02)
    \log_{10}(\mathcal{L}_{{}_{5 \, \mbox{\tiny GHz}}}/\mbox{erg} \cdot
    \mbox{s}^{-1}).
    \label{eq16}
  \end{equation}

  With formulas (\ref{eq11}) - (\ref{eq16}), the GLF can be already calculated. 
  The limits on $z$ and $\mathcal{L}_\gamma$ defined for the gamma-ray
  luminosity function in \cite{GLF-inoue11}  are summarized in table
  \ref{tab3}. These values are used to define the integration bounds of
  expression (\ref{eq4}) in the case of FRI galaxies. The limits on
  $\epsilon_\gamma$ in $\mathcal{L}_\gamma$ are also presented in table
  \ref{tab3}. Accordingly, they were used to evaluate
  $\mathcal{L}^{\mbox{\tiny{fit}}}_\gamma$ in equation (\ref{eq10}). For
  Centaurus A (M87), $\mathcal{L}^{\mbox{\tiny{fit}}}_\gamma(\epsilon_{\gamma_1},
  \epsilon_{\gamma_2}) = 6.3 \times 10^{40} \, (4.4 \times 10^{41}) \,
  \mbox{erg} \cdot \mbox{s}^{-1}$.

  \subsection{The GLF for BL Lac galaxies}

   For BL Lac galaxies the gamma-ray luminosity function (GLF) is taken from 
   reference \cite{Fermi-list}. There, the GLF is calculated on the basis of a 
   sample of 42 BL Lac galaxies detected with the \textit{Fermi}-Lat telescope
   during its first three months of sky-survey at galactic latitudes $|b| >
   10^\circ$. The GLF derived by the \textit{Fermi}-LAT  Collaboration for BL
   Lac galaxies per interval of comoving volume and $\log_{10}\mathcal{L}_\gamma$ 
   is presented below:
   \begin{equation}
     \rho_\gamma(\mathcal{L}_\gamma, z) =  
      \frac{\rho_B}{\log_{10} e}
      \left( \frac{\mathcal{L}_\gamma}{10^{48} \, \mbox{erg} \cdot
      \mbox{s}^{-1}} \right)^{-\zeta + 1}.
     \label{eq17}
   \end{equation}
   Here, $\rho_\gamma$ is measured in $\mbox{cm}^{-3}$. $\mathcal{L}_\gamma$ is 
   the integrated gamma-ray luminosity for $\epsilon_\gamma >100 \,
   \mbox{MeV}$. The values of the parameters $\rho_B$ and $\zeta$ are
   estimated in \cite{Fermi-list} for two different bins of redshift: $z = 0.0
   - 0.3$ and $z > 0.3$. The respective values are shown in table \ref{tab4}.

 \begin{table}[!t]
 \begin{center}
 \caption{Limits on $z$ and $\mathcal{L}_\gamma$ for the gamma-ray luminosity
 functions of FRI \cite{GLF-inoue11} and BL Lac galaxies \cite{Fermi-list}, and
 definitions of the $\epsilon_{\gamma}$ integration bounds for $\mathcal{L}_\gamma$.}
 \begin{tabular}{l|cccccc}
 \hline
  Galaxy & $z_{min}$ & $z_{max}$ & $\mathcal{L}_{\gamma}^{min} [\mbox{erg} \cdot \mbox{s}^{-1}]$ &
  $\mathcal{L}_{\gamma}^{max} [\mbox{erg} \cdot \mbox{s}^{-1}]$ &
  $\epsilon_{\gamma_1} [\mbox{TeV}]$ & $\epsilon_{\gamma_2} [\mbox{TeV}]$\\
 \hline
  FRI    &0.0&5.0& $1 \times 10^{39}$& $1 \times 10^{48}$ & $10^{-4}$ & $10^{-2}$\\
  BL Lac &0.0&0.3& $2 \times 10^{44}$& $1 \times 10^{46}$ & $10^{-4}$ & $10^{5}$\\
  BL Lac &0.3&1.2& $2 \times 10^{46}$& $4 \times 10^{48}$ & $10^{-4}$ & $10^{5}$\\
 \hline
 \end{tabular}
 \label{tab3}
 \end{center}
 \end{table}

   The integration limits for equation (\ref{eq4}) are summarized in table
   \ref{tab3}. They are chosen to cover roughly the luminosity and redshift 
   ranges of the BL Lac sample reported in \cite{Fermi-list}.  

 \begin{table}[!b]
 \begin{center}
 \caption{Values of the parameters of the GLF function (\ref{eq17}) for BL Lac 
 galaxies and two different redshift bins \cite{Fermi-list}.}
 \begin{tabular}{l|ccccc}
 \hline
  Redshift bin & & $\zeta$ & & $\rho_B [10^{-85} \, \mbox{cm}^{-3}]$\\
 \hline
  $z = 0.0 - 0.3$& & 2.08    & & 2.61\\
  $z > 0.3$      & & 2.10    & & 1.62\\
 \hline
 \end{tabular}
 \label{tab4}
 \end{center}
 \end{table}

 \section{The gas and dust content of individual sources}

  \subsection{Centaurus A}

  \subsubsection{Model for FRI galaxies based on CenA}

\begin{figure}[!t]
 \centering
 \includegraphics[width=3.2in]{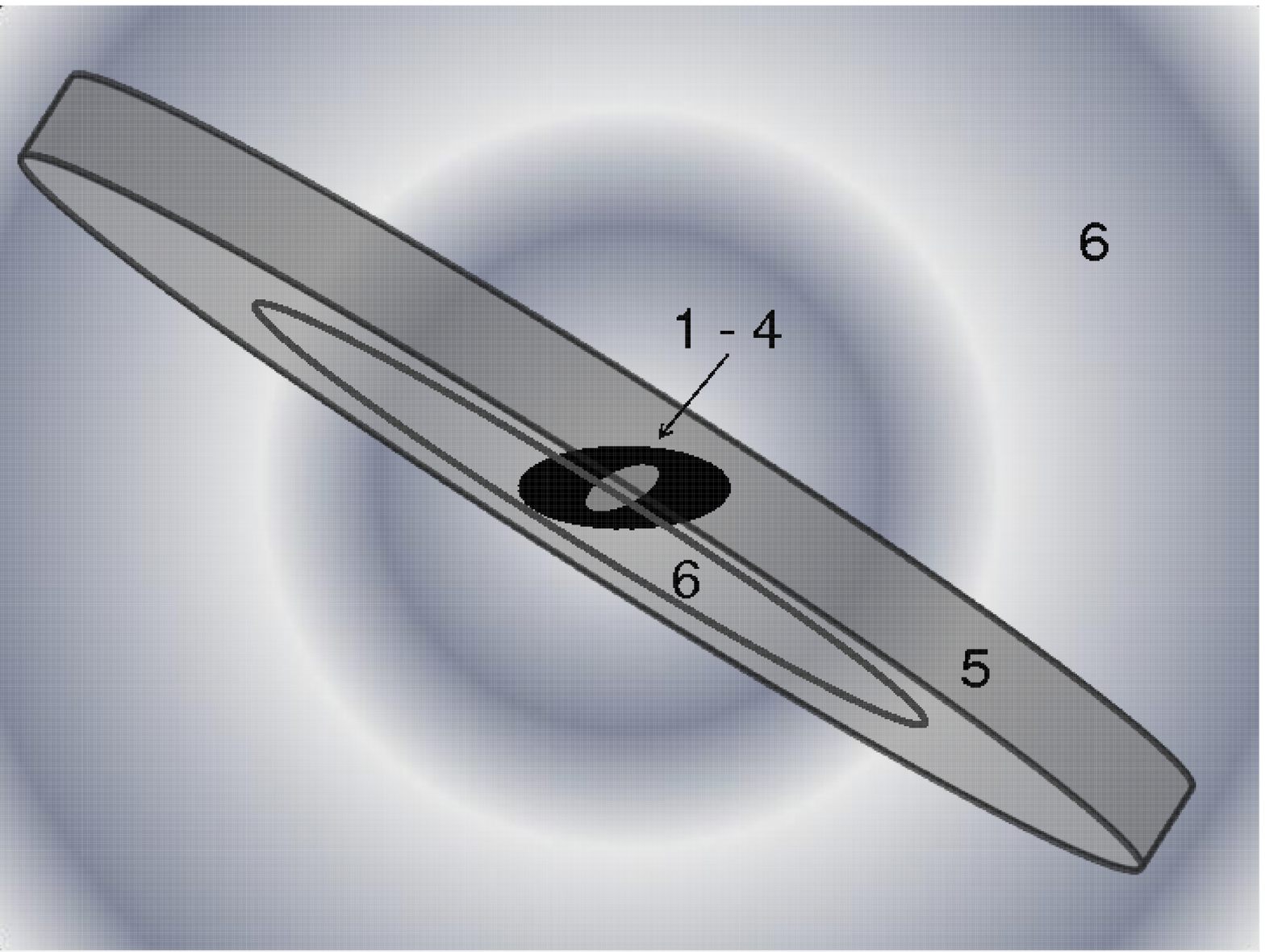}
 \includegraphics[width=3.2in]{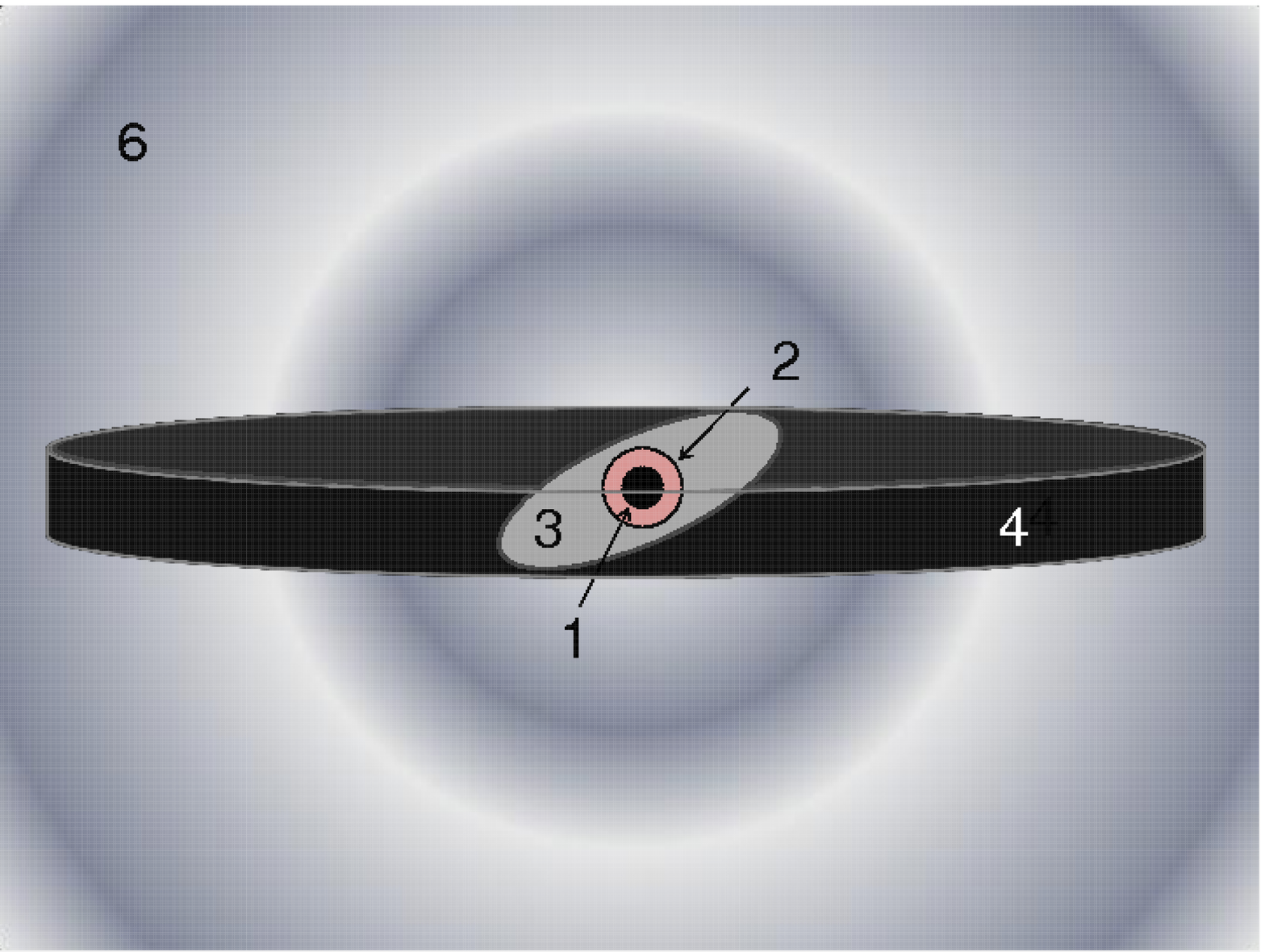}
  \caption{Model \textbf{CenA1} for the matter distribution in Centaurus A. In
  the figures, the following structures are represented: (1) core, (2) X-ray
  absorber, (3) nuclear disk, (4) circumnuclear disk, (5) dust lane and (6)
  ISM and halo. On the left, the biggest structures of the galaxy are
  shown. On the right, a close-up of the inner region of  Centaurus A is
  presented.}
 \label{fig2}
\end{figure}

  Due to its proximity to Earth, Centaurus A has been well studied in the
  literature (see for example, the reviews \cite{Morganti10, Israel98}). 
  To calculate the angle-averaged gas column density,  $\bar{\Sigma}_{\small
  H}$, traversed by the gamma-ray flux at source, a simple model (called
  \textbf{CenA1}) is built based on the observational data. A sketch depicting
  the main structures of the model is shown in figure \ref{fig2}. The main
  parameters are presented in table \ref{tab5}. 
  
  The most prominent features of Centaurus A are the nuclear region, the
  circumnuclear disk, the dust lane, the jets and the halo \cite{Morganti10, 
  Israel98}. All of them,  but the jet structures, are included in the
  model. Accordingly, the corresponding $\bar{\Sigma}_{\small H}$ will be 
  used only for FRI radio galaxies, where the jets are not aligned with the
  line of sight.

  In this model of Centaurus A, the VLBI radio core \cite{Kellerman97} it is
  assumed to be the source of the $\gamma$ radiation. This supposition is not yet
  confirmed, but it is still consistent with observations. Actually, the
  observed position of the gamma-ray core encompasses the radio core
  and the inner kpc jets \cite{centA-hess09, centA-fermi10}). An upper limit 
  of $R = 0.01 \, \mbox{pc} = 3.09 \times 10^{16} \, \mbox{cm}$ has been
  established for the radius of the source from VLBI observations
  \cite{Kellerman97}. But according to some models based on observational data
  \cite{centA-fermi10, Abraham07} the actual size could be smaller, of
  the order of $10^{15} \, \mbox{cm}$. Here, it will be taken $R_{1} = 0.01 \,
  \mbox{pc}$, which is measured from the center of the nucleus of the
  galaxy. For the gas density, the situation is also uncertain. Assuming
  free-free absorption in the core of Centaurus A and using data from mm and
  X-ray observations, in \cite{Abraham07} a rough estimation of $n_{\small H}
  \approx 10^6 \, \mbox{cm}^{-3}$ is given, value that will be used for this
  work.

  X-ray observations of the nucleus of Centaurus A reveals the presence of two
  spectra with different degrees of nuclear absorption \cite{Suzaku07}. The
  spectra could be associated either with the emission of a single X-ray 
  source attenuated by different components of a strong absorber \cite{Suzaku07,
  Turner97, Wozniak98} or  with the combined emission from the accretion disk
  and the pc-scale VLBI jet \cite{centA-Evans04}. In this paper, the first
  scenario will be used to get insight into the nuclear environment of
  Centaurus A. Here, it will be assumed that the absorber is in the form of a
  cold cloud entirely surrounding the core \cite{Miyazaki96} characterized by
  a column density with weighted-mean of $2.4 \times 10^{23} \,
  \mbox{cm}^{-2}$ (best-fit model from \cite{Suzaku07}). The cloud is supposed 
  to have a uniform density ($n_{\small H} \approx 8.64 \times 10^5 \,
  \mbox{cm}^{-3}$) and a spherical geometry with radius $R_{2} = 0.1 \,
  \mbox{pc}$, which is taken from the estimated emission radius of the Fe
  K$\alpha$ lines observed in the X-ray spectra, presumably belonging to the
  absorber \cite{centA-Evans04}. Although, such a cloud may not be
  representative of the environments of the AGN (see, for example,
  discussions in \cite{centA-Evans04, Suzaku07, Wozniak98}), 
  it is a simple model to deal with that in some degree reproduces the 
  observed column depths.

  SINFONI observations of the central region in Centaurus A reveals that the 
  corresponding gas morphology is consistent with a warped-disk structure formed by 
  several inclined rings \cite{Neumayer07}. Accordingly, the region between 
  $R_{2} = 0.1 \, \mbox{pc}$ and $R_{3} = 40 \, \mbox{pc}$ will be described using 
  the warped-disk model \cite{Quillen06, Tubbs80}. The position ($PA$) and 
  inclination ($i$) angles\footnote{Here, following reference \cite{Quillen09}, 
  the $PA$ is just the angular orientation in the sky of the major axis of the 
  galaxy and it is measured counter clockwise from north to the blue shifted side 
  of the aforementioned axis. On the other hand, $i$ is the angle of 
  inclination of the galaxy to the line of sight. For an edge-on galaxy,
  $i$ is equal to $90^\circ$. Meanwhile, for a galaxy with its northern (southern) side 
  closer to the observer, the value of $i$ lies above (below) $90^\circ$.} 
  for the tilted rings of the model are taken from \cite{Neumayer07}
  and are presented in table \ref{tab6}. Since data is reported only between
  $0.824 - 31.3 \, \mbox{pc}$ \cite{Neumayer07}, the values of the disk parameters 
  outside this range, but inside $R=0.1$ and $40 \, \mbox{pc}$, will be kept
  constant. For simplicity, a disk aspect ratio  $k(R) = h(R)/R = 0.5$ is
  assumed, where $h(R)$ is the thickness as a function of the cylindrical
  radius, $R$.

\begin{table}[!t]
 \begin{center}
 \caption{Extension and density distributions of the main gas structures 
  in the model \textbf{CenA1}. Here, $R$ represents
  the cylindrical radius and $r$, the spherical one.}
 \begin{tabular}{lccc}
 \hline
  Structure              & & $R\,[\mbox{pc}]$         & $n_{\small H} \, [\mbox{cm}^{-3}]$ \\
 \hline
 Core                    & & $0.01$                   & $10^{6}$\\
 X-ray absorber          & & $0.01 - 0.1$             & $8.6 \times 10^{5}$\\
 Nuclear disk            & & $0.1  - 40 $             & $3.4 \times 10^{2}$\\
 Circumnuclear disk      & & $40 - 200  $             & $(5.5 \times 10^{5})/r^2$\\
 Dust lane               & & $(0.8 - 7)\times 10^{3}$ & $(3.6 \times 10^{13})/r^4$\\
 ISM and halo            & & $35 \times 10^{3}$       & $4 \times 10^{-2}  [1 + (r/500 \, \mbox{pc})^2]^{-0.6}$\\
 \hline
 \end{tabular}
 \label{tab5}
 \end{center}
 \end{table}

  The next important structure of Centaurus A is the circumnuclear disk \cite{Morganti10, 
  Israel98}. Through independent $CO(1-0)$ absorption measurements against 
  the nucleus \cite{Israel90} and detection of $CO(2-1)$ emission lines \cite{Espada09}, 
  it was found that the gas distribution agrees with a disk-like feature with 
  $PA = 155^\circ$ and $i = 70^\circ$, external radius $R_4 = 200 \, \mbox{pc}$,  
  thickness of $80 \, \mbox{pc}$ and gas mass of $\sim 10^{7} M_\odot$. The disk has 
  an inner cavity of radius $R = 40 \, \mbox{pc}$ as suggested by $H_2$
  emission studies towards the core of the galaxy \cite{Israel90}. Following
  reference \cite{Israel90}, a $n_{\small H} \sim r^{-2}$ density distribution
  will be assigned to the circumnuclear disk, where $r$ is the spherical
  radius. To estimate the density inside the nuclear disk, the above density
  function is evaluated using $R = 40  \, \mbox{pc}$.

  A gap in the gas distribution is present between $R = 200 - 800 \, \mbox{pc}$  
  as suggested by several studies \cite{Quillen06, Espada09, Marconi00}. Beyond 
  this region, the dust lane is found \cite{Morganti10, Israel98}. A warped-disk
  model with inclined tilted-rings \cite{Nicholson92, Quillen92} seems to describe 
  the optical, $H\alpha$, $CO$ and infrared  observations of the dusty disk
  (see \cite{Morganti10} and references therein). The radius of the dust lane is 
  about $R_5 = 7 \, \mbox{kpc}$ and its mass is of the order of $1.3 \times 10^{9} 
  M_\odot$ \cite{Israel98}. To describe the disk aspect ratio, the model of
  reference \cite{Quillen06} is adopted: $k(R) = h(R)/R = 0.1 (R/824 \, \mbox{pc})^{0.9}$.
  The values for the position and inclination angles of the tilted-rings in the dust
  lane are presented in table \ref{tab6}. They are taken from references 
  \cite{Quillen06, Espada09, Quillen09}. A simple $r^{-4}$ power-law dependence 
  will be assumed for the density distribution of the dusty disk
  \cite{Quillen06, Quillen92}.

 \begin{table}[!t]
 \begin{center}
 \caption{Position and inclination angles for the nuclear, circumnuclear
  and dusty disks. Data are taken from the compilation shown in \cite{Quillen09}.
  The original references are also shown. Extrapolations from the measured data
  have been applied for the intervals $r = 0.10 - 0.82, 31.3-40 \, \mbox{pc}$ and 
  $r = 5.69 - 7 \, \mbox{kpc}$.}
 \begin{tabular}{ccc|ccc|ccc}
 \hline
 \multicolumn{9}{c}{Nuclear disk \cite{Neumayer07}}\\
 \hline
  $R\,(\mbox{pc})$ & $i\,({}^\circ)$ & $PA\,({}^\circ)$&
  $R\,(\mbox{pc})$ & $i\,({}^\circ)$ & $PA\,({}^\circ)$&
  $R\,(\mbox{pc})$ & $i\,({}^\circ)$ & $PA\,({}^\circ)$\\
 \hline  
 $0.10 - 0.82$& 45.0&   144.0& $8.41$&  44.0&   168.9& $19.5$&  39.7&   149.3\\
 $0.82$&        45.0&   144.0& $9.40$&  39.5&   169.0& $20.8$&  42.5&   147.8\\
 $1.65$&        37.6&   148.5& $10.4$&  36.1&   167.2& $22.1$&  46.7&   146.8\\
 $2.64$&        38.8&   152.1& $11.4$&  35.1&   165.2& $23.4$&  48.9&   145.1\\
 $3.63$&        42.2&   154.8& $12.5$&  34.2&   163.5& $24.9$&  51.6&   144.0\\
 $4.45$&        44.5&   160.5& $13.5$&  35.2&   161.7& $26.4$&  53.6&   142.8\\
 $5.44$&        42.3&   165.4& $14.7$&  36.9&   159.8& $27.9$&  55.6&   141.6\\ 
 $6.43$&        44.1&   167.8& $15.8$&  36.8&   157.0& $29.5$&  54.6&   137.9\\
 $7.42$&        45.7&   168.5& $17.0$&  36.2&   153.8& $31.3$&  58.9&   134.9\\
       &            &        & $18.1$&  36.8&   151.1& $31.3-40$&  58.9&   134.9\\
 \hline
 \multicolumn{9}{c}{Circumnnuclear and dusty disks \cite{Espada09}}\\
 \hline
  $R\,(\mbox{pc})$  & $i\,({}^\circ)$ & $PA\,({}^\circ)$&
  $R\,(\mbox{kpc})$ & $i\,({}^\circ)$ & $PA\,({}^\circ)$&
  $R\,(\mbox{kpc})$ & $i\,({}^\circ)$ & $PA\,({}^\circ)$\\
 \hline
 $40 - 200$&    70.0  &  155.0 & $0.80$&        85&     94 & $1.09$&    84&     118\\
           &          &        & $0.84$&        84&     97 & $1.14$&    85&     121\\
           &          &        & $0.89$&        82&     102& $1.19$&    88&     124\\ 
           &          &        & $0.94$&        82&     107& $1.24$&    90&     127\\ 
           &          &        & $0.99$&        82&     111& $1.29$&    93&     128\\
           &          &        & $1.04$&        83&     115& $1.34$&    96&     130\\
 &&&&&&$1.38$&  100&    131\\
 \hline
 \multicolumn{9}{c}{Dusty disk \cite{Quillen06}}\\
 \hline
  $R\,(\mbox{kpc})$ & $i\,({}^\circ)$ & $PA\,({}^\circ)$&
  $R\,(\mbox{kpc})$ & $i\,({}^\circ)$ & $PA\,({}^\circ)$&
  $R\,(\mbox{kpc})$ & $i\,({}^\circ)$ & $PA\,({}^\circ)$\\
 \hline
 $1.40$&        104.8&  132 & $1.81$&   118.5&  125 & $2.23$&   110.2&  127\\
 $1.48$&        110.3&  131 & $1.90$&   117.9&  125 & $2.31$&   107.6&  127\\
 $1.57$&        114.5&  129 & $1.98$&   116.7&  125 & $2.39$&   105  &  127\\
 $1.65$&        117.1&  127 & $2.06$&   114.9&  126 & $2.47$&   102.6&  126\\
 $1.73$&        118.3&  126 & $2.14$&   112.7&  127 &       &        &     \\
 \hline
 \multicolumn{9}{c}{Dusty disk \cite{Quillen09}}\\
 \hline
  $R\,(\mbox{kpc})$ & $i\,({}^\circ)$ & $PA\,({}^\circ)$&
  $R\,(\mbox{kpc})$ & $i\,({}^\circ)$ & $PA\,({}^\circ)$&
  $R\,(\mbox{kpc})$ & $i\,({}^\circ)$ & $PA\,({}^\circ)$\\
 \hline
  $2.72$&       105.5&  126.9 & $3.21$& 94.2&   117.4 & $4.86$& 90&     105\\
  $2.97$&       99.6 &  125   & $4.04$& 90  &   110.3 & $5.69$& 90&     102\\
  &&&&&&$5.69 - 7$&     90&     102\\
 \hline
 \end{tabular}
 \label{tab6}
 \end{center}
 \end{table}

  Finally,  XMM-Newton observations of the hot ISM of Centaurus A in X-rays 
  have revealed the presence of a large-scale gas halo extending at least up to
  $R_6 = 35 \, \mbox{kpc}$ \cite{Kraft09}. An  average gas density was determined 
  in \cite{Kraft03} from fits to Chandra and XMM-Newton observations, where it was shown 
  that for $r \lesssim 10 \, \mbox{kpc}$, it is well described by a beta model 
  $n_{\small H} = n_0  [1 + (r/r_0)^{-1.5\xi}]$ with $n_0 = 4 \times 10^{-2} \, 
  \mbox{cm}^{-3}$, $r_0 = 500 \, \mbox{pc}$ and  $\xi = 0.40 \pm 0.04$. This
  profile will be extrapolated up to $r = 35 \, \mbox{kpc}$ \cite{Kraft09}. 
  Additionally, it will be used to fill out the gaps among the internal
  structures.

  Besides the above model, a second one, named \textbf{CenA2}, will be also used, 
  in which the nuclear region is left out. The aim is to include the extreme
  case in which the parent particles (lepton or hadrons) involved in the
  $\nu-$production chain are completely absorbed in the nuclear zone. For this
  model, the estimation of the column depth is performed by integrating the
  proton density for the ISM and Halo inside the interval $r = 40 - 35 \times
  10^3 \, \mbox{pc}$. The corresponding value will be considered as a lower
  value for $\bar{\Sigma}_{\small H}$.\\

  \subsubsection{Model for BL Lac galaxies based on CenA}

  Since for this type of galaxies the jet lies along the line of sight of the observer,
  interactions of the gamma-rays with the matter of the jet should be considered.
  For this calculation, the target of the $\gamma$ radiation will be a
  homogeneous and spherical distribution of relativistic gas flowing outwards
  (with $\Gamma  = 10$) and composed of protons (model \textbf{JetCenA}). The
  radius and density of the gas cloud are set equal to the extension and
  average hydrogen density of the jet of Centaurus A, respectively. For the
  former, $r \sim  4 \, \mbox{kpc}$ \cite{Kraft02} and for the latter,
  $n_{\small H} = 1.7 \, \mbox{cm}^{-3}$ \cite{Kachelriess09}, both values were
  derived from X-ray observations performed with the Chandra telescope.

   \begin{figure}[!t]
   \centering
    \includegraphics[width=3.2in]{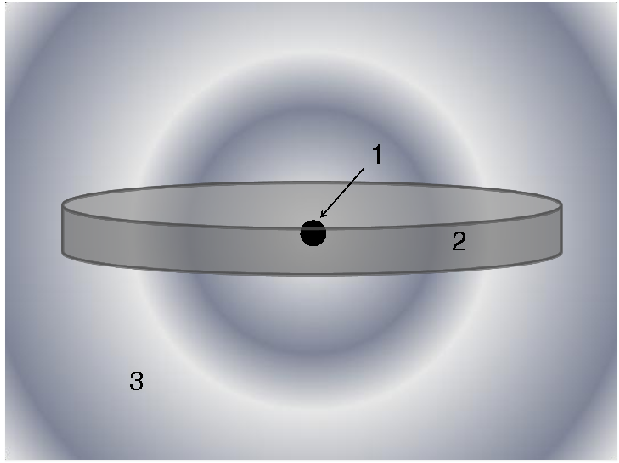}
    \caption{Model \textbf{M87a} for the matter distribution of the radio
    galaxy M87. In the figure, the following structures are represented
    schematically: (1) core, (2) nuclear disk, and (3) ISM and halo.}
   \label{fig3}
  \end{figure}

  \subsection{M87}

  \subsubsection{Model for FRI galaxies based on M87}

   M87 is one of the closest radio galaxies to Earth, therefore it has become
   the target of several astronomical observations dedicated to get a deeper 
   understanding of the astrophysics and structure of this type of galaxies.  
   Observations of M87 reveal the presence of a core \cite{Cohen69}, a nuclear 
   disk  \cite{Ford94}, a kpc jet \cite{Curtis18, Biretta95} and a halo 
   \cite{Malina76, Schreier82, Sarazin88}. To built the model for M87, as it was the case 
   for Centaurus A,  the contribution of the jet will be neglected.

   The source of gamma rays in this model (\textbf{M87a}), will be located at the core 
   of the radio galaxy in accordance with the results of the $2008$ multi-wavelength 
   on M87 \cite{m87-multi09}. The results show evidence in favor of a
   correlation between the radio core and the TeV emissions. Assuming that the
   first source is located in the central region of the galaxy, it is
   concluded that the  gamma radiation should be produced in the immediate
   vicinity of the black hole \cite{m87-multi09}. Following the model
   proposed in \cite{m87-fermi09} to explain the \textit{Fermi}-LAT MeV/GeV
   observations on the basis of synchrotron self-Compton emissions from
   electrons at the core, a gamma source with a radius of $r_1 = 1.4 \times
   10^{16} \, \mbox{cm} = 4.5 \, \times 10^{-3} \, \mbox{pc}$ is assumed. This value is
   consistent with both the limits derived from VLBA observations at 43
   GHz \cite{Junor99, Ly07} and the constraints obtained with HESS
   \cite{Aharonian06}, MAGIC \cite{Albert08} and VERITAS \cite{Acciari10} from
   the observed variability (of the order of a few days) of the TeV
   radiation. The density of the source is uncertain. Here, the value  
   $n_{\small H} = 10^6 \, \mbox{cm}^{-3}$ will be used, which is in agreement 
   with the limits presented in \cite{Neronov07} for the plasma density at 
   the nucleus of M87 and with estimations derived in \cite{Sabra03} from the nuclear
   emission lines measured with the instruments on board of the Hubble Space 
   Telescope (\textit{HST}) in the optical and UV regions, when data is 
   interpreted within the framework of photoionization scenarios.

   M87 has a nuclear disk of ionized gas and radius $r \sim 100 \, \mbox{pc}$
   rotating around a massive central object as revealed by the \textit{HST} 
   \cite{Ford94, Ford98}. The disk has a spiral structure, which seems 
   to be connected with large-scale filaments of gas ($\sim 1 \, 
   \mbox{kpc}$ long) \cite{Ford94, Ford98}. These warped structures 
   will not be considered in the present work. The major axis position angle
   and the inclination of the disk are $PA = 6^\circ$ and $i = 35^\circ$, 
   respectively \cite{Tsve98}. Regarding the mean surface density of the disk,
   a rough estimation of $\Sigma_{\small H} = 2.5 \times 10^{19} \, \mbox{cm}^{-2}$ 
   was obtained in \cite{Dopita97} within the central $70 \, \mbox{pc}$ using data 
   from the HST \cite{Ford94}. In reference \cite{Sabra03}, the total column
   density of neutral hydrogen in the disk was constrained to the interval
   $\Sigma_{\small H} = 10^{19} - 10^{20} \, \mbox{cm}^{-2}$ by combining
   optical, UV, X-ray and HI 21 cm observations of the nucleus of M87. 
   The presence of cold molecular gas ($H_2$) could largely increase the
   column density of gas. From observations of the nucleus of M87 with the 
   Submillimeter Array an upper limit $\Sigma_{\small H_2} \leq 4.5  \times
   10^{22} \, \mbox{cm}^{-2}$ was found in \cite{Tan08} within $\sim 100 \,
   \mbox{pc}$. For the present work, a mean column density of $\Sigma_{\small
   H} = 10^{20} \, \mbox{cm}^{-2}$ in a plane perpendicular to the disk will
   be assumed. The nuclear disk will be  modeled using a thin disk with radius
   $r = 100 \, \mbox{pc}$ and thickness of $10 \, \mbox{pc}$ as in reference
   \cite{Tan08}.

 \begin{table}[!t]
 \begin{center}
 \caption{Extension and density distributions of the main gas structures 
  in the model \textbf{M87a}. Here, $R$ represents
  the cylindrical radius and $r$, the spherical one.}
 \begin{tabular}{lccc}
 \hline
  Structure              & & $R\,[\mbox{pc}]$         & $n_{\small H} \, [\mbox{cm}^{-3}]$ \\
 \hline
 Core                    & & $0.0045$                 & $10^{6}$\\
 Nuclear disk            & & $100$                    & $3.24$\\
 Halo                    & & $260 \times 10^{3}$      & $4.2 \times 10^{-2}  [1 + (r/7.87 \, \mbox{kpc})^2]^{-0.654}$\\
 \hline
 \end{tabular}
 \label{tab7}
 \end{center}
 \end{table}   

   The mass and matter distribution of the hot gas in the halo of M87 has been well 
   studied in the literature in the X-ray window with several space observatories, 
   for example, Einstein (see \cite{Sarazin88} and references therein), 
   ROSAT \cite{Nulsen95} and XMM-Newton \cite{Boehringer01}. In reference \cite{Fabricant83},
   using X-ray measurements obtained with the Einstein observatory, the density profile 
   of the ISM in the halo of M87 was derived up to a distance of $392 \, \mbox{kpc}$ 
   from the nucleus. The distribution was estimated assuming a gas in hydrostatic equilibrium
   with a spherical distribution described by a beta model  $n_{\small H} = n_0  
   [1 + (r/r_0)^{-1.5\xi}]$. The corresponding analysis gave the following parameter
   values: $n_0 = 4.2 \times 10^{-2} \, \mbox{cm}^{-3}$, $r_0 = (7.87 \pm 1.36) \times 10^{3} \, 
   \mbox{pc}$ and  $\xi = 0.436 \pm 0.008$. The above formula will be 
   incorporated to the present model of M87 in order to describe the matter
   distribution of the halo up to  $r = 260 \, \mbox{kpc}$, where the measured gas is still
   associated with M87 \cite{Fabricant83}. A figure, showing  the main components 
   of M87 here discussed, is presented in figure \ref{fig3}. The parameters
   of the model are shown in Table \ref{tab7}.\\

   Another model, called \textbf{M87b},  will be included into the
   calculations of the neutrino flux, which does not involve the nuclear
   region. The column depth in this case is calculated solely from the beta
   model for the ISM and halo density, from $r = 40$ to $260 \,
   \mbox{kpc}$. This model is introduced to derive a lower limit on
   the column density of target protons for an extreme scenario where the
   parent particles of neutrinos are completely absorbed before decaying in
   the internal regions of the AGN. \\

  \subsubsection{Model for BL Lac galaxies based on M87}

  Proceeding similarly to the case of Centaurus A, a spherical cloud of gas
  with homogeneous density and composed of protons will be considered (model
  \textbf{JetM87}). The gas is flowing outwards with Lorentz factor $\Gamma =
  10$. The radius of the cloud will be  $r = 2 \, \mbox{kpc}$, which is the
  length of the jet of the giant radio galaxy M87 \cite{Biretta95}, and the
  density, $n_{\small H} \approx 1 \, \mbox{cm}^{-3}$, as in reference
  \cite{Reynolds96}. To end this section the angle-averaged column densities of 
  protons estimated for the models already presented are summarized in table
  \ref{tab8}.

 \begin{table}[!t]
 \begin{center}
 \caption{Column depths calculated for Centaurus A and M87 using different
  models (see text). \textbf{JetCenA} and   \textbf{JetM87} will be employed
  for the description BL Lac galaxies, while the rest of the models, for 
  FRI objects.}
 \begin{tabular}{lccccccc}
 \hline
  & & CenA1  & CenA2 & JetCenA & M87a & M87b & JetM87\\
  $\bar{\Sigma}_{\small H} (10^{21} \,\mbox{cm}^{-2})$  & & $4.58 \times 10^{2}$ & $0.21$  & $20.98$ & 
  $16.93$ & $2.79$ & $6.17$\\
 \hline
 
 \hline
 \end{tabular}
 \label{tab8}
 \end{center}
 \end{table}

  \section{The $\gamma P$ hadronic cross section and the neutrino yields}

  For the calculation of the total $\gamma P$ hadronic cross section above
  $\sqrt{s} = 5 \, \mbox{GeV}$  the following expression is employed
  \cite{PDG10, Compete02}:
  \begin{equation}
   \sigma_{\gamma P} = \delta\cdot Z +  \delta\cdot B \log^2(s/s_0) + Y(s_1/s)^{\eta_1}.
   \label{eq18}
  \end{equation} 
  where  $\sqrt{s}$ is defined as the center-of-mass energy and $\sqrt{s_1}$ is
  fixed at $1  \, \mbox{GeV}$. On the other hand, $\delta = 0.00308$, $Z = 35.45 \,
  \mbox{mb}$, $B = 0.308 \, \mbox{mb}$,  $Y = 0.0320 \, \mbox{mb}$ and $\eta_1
  = 0.458$, while $\sqrt{s_0} = 5.38 \ \mbox{GeV}$ \cite{PDG10}. These
  parameters were obtained by the COMPETE Collaboration by means of a global
  fit to current accelerator data with $\sqrt{s} \geq 5  \, \mbox{GeV}$
  \cite{Compete02}, where the multi-pion production channel is the dominant
  one. Since, formula (\ref{eq18}) can not be applied below $\sqrt{s} = 5  \,
  \mbox{GeV}$, interpolation techniques will be applied on the total cross
  section data presented in \cite{PDG10} to get $\sigma_{\gamma P}$
  up to $1 \, \mbox{GeV}$. It is in this low-energy regime, where baryon
  resonances and direct-pion production processes become also relevant
  \cite{Sophia, Hummer10}.

 \begin{figure}[!t]
 \centering
 \includegraphics[width=3.2in]{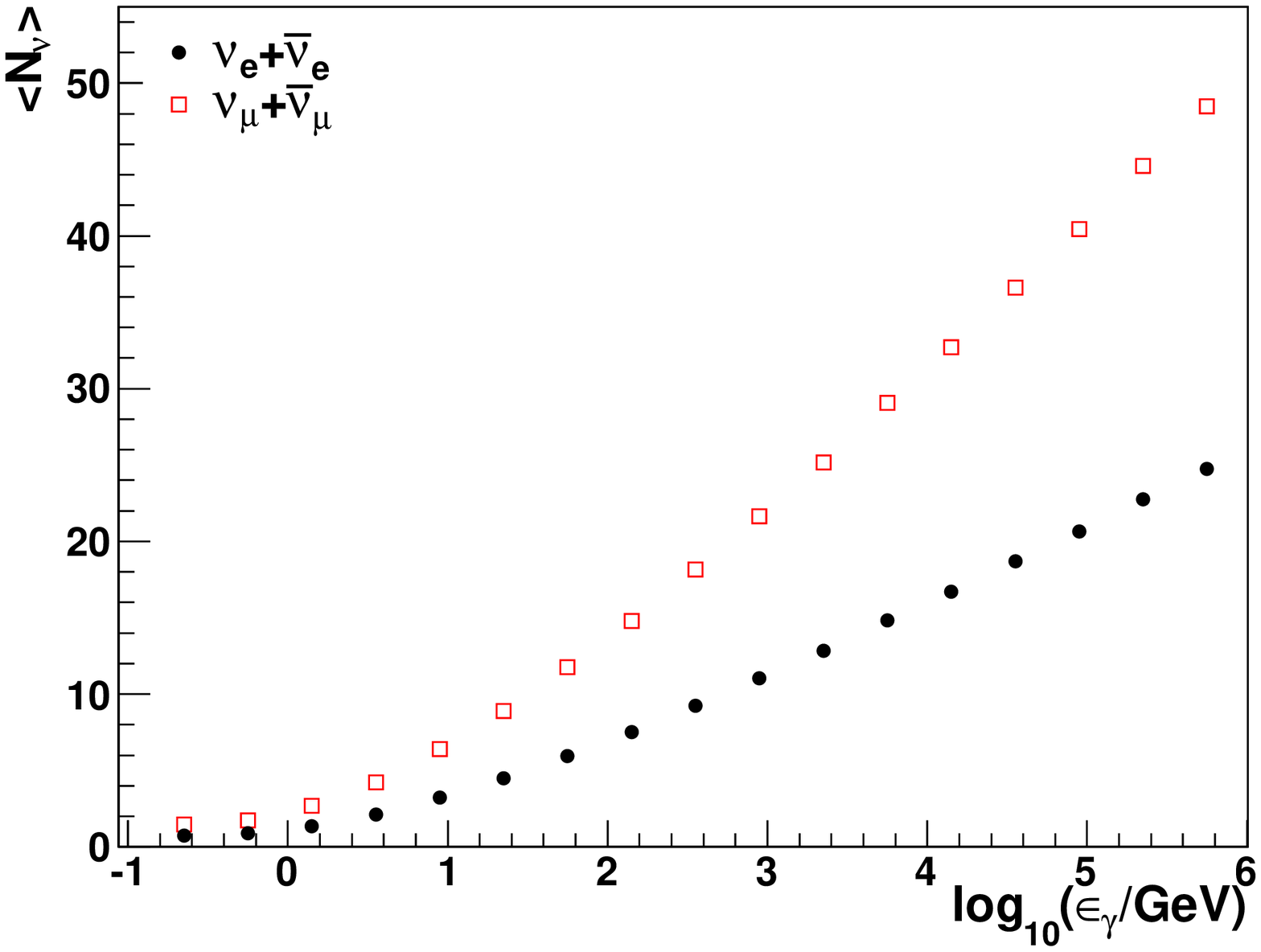}
 \includegraphics[width=3.2in]{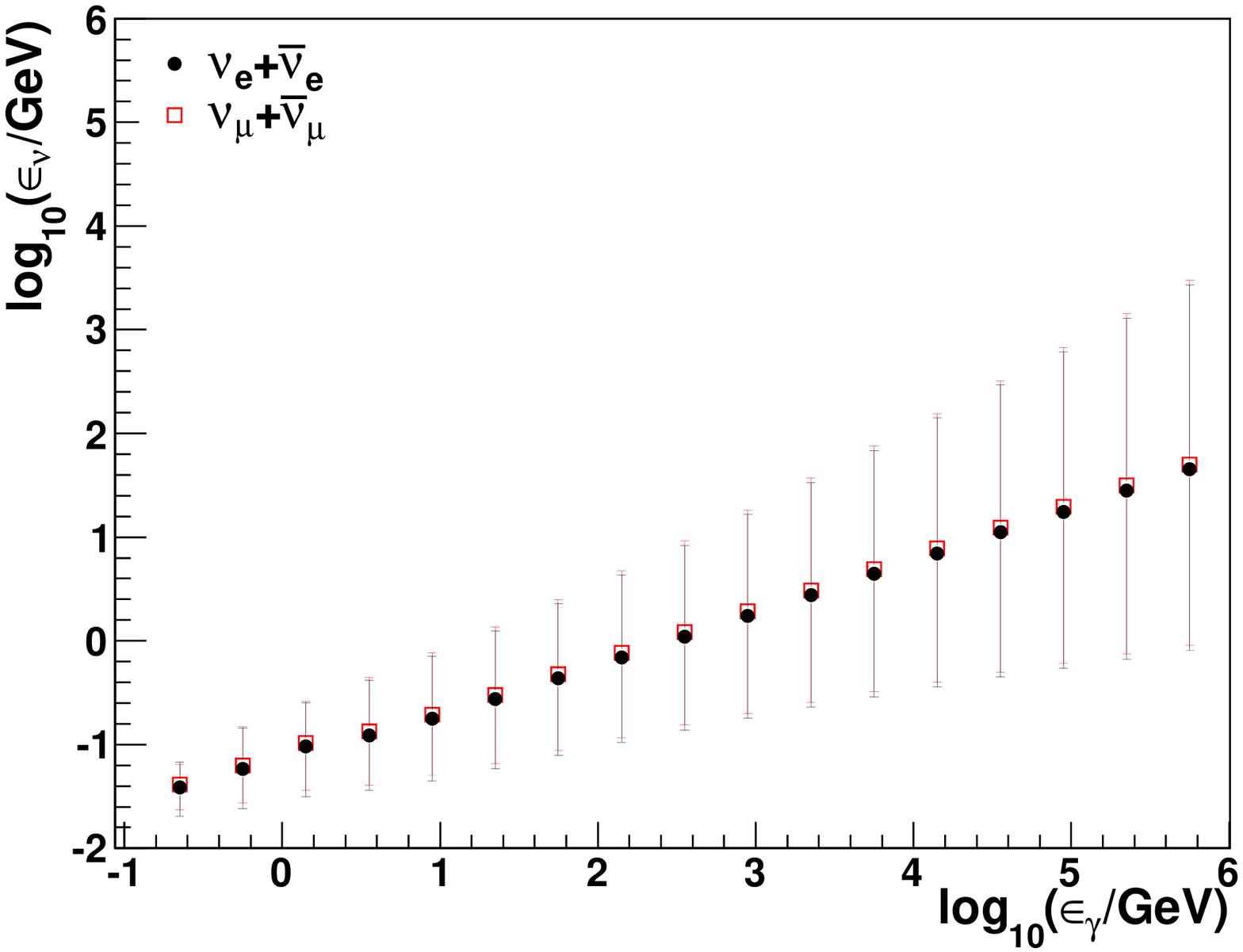}
  \caption{Left: Average electron (muon) neutrino and antineutrino multiplicities for
  $\gamma$P collisions. Right: Median of the neutrino and antineutrino energy distributions 
  for $\gamma$P reactions. Error bars represent the $84\%$ and $16\%$ quantiles.
  $\epsilon_\gamma$ is the energy of the incident photon.}
 \label{fig4}
\end{figure}

  The neutrino yields are obtained using the Monte Carlo program SOPHIA v2.01
  \cite{Sophia}. Estimates are performed only for electron and muon neutrinos,
  since these neutrino flavors are more copiously produced than tau neutrinos
  in $\gamma P$ interactions in the energy regime considered in this work. 
  To obtain the neutrino yield $Y^{\gamma P\rightarrow \nu_\ell}(\epsilon_\gamma,
  \epsilon_\nu)$,  a 2D histogram is built. Horizontal and vertical axes
  correspond to the logarithm of the photon and neutrino energies, respectively.
  The energy intervals involved are $\log_{10}(\epsilon_\nu/\mbox{GeV}) = [-5,6]$ 
  and $\log_{10}(\epsilon_\gamma/\mbox{GeV}) = [-0.8,6]$, while the size of
  the bins is to $\Delta \log_{10}(\epsilon_\nu/\mbox{GeV}) = 0.1$.
   
  The bin $(i, j)$ is filled with the average number of neutrinos, $\nu_\ell$,
  and antineutrinos, $\bar{\nu}_\ell$, with energy in the interval $\Delta
  \log_{10}(\epsilon_{\nu, j})$ produced per $\gamma P$ interaction at photon 
  energies $\epsilon_{\gamma, i}$ in the laboratory frame. Protons are
  considered to be at rest in the same reference system. Finally, for each
  photon energy, $\epsilon_{\gamma, i}$, around $10^5$ events are simulated. 

  The mean multiplicity and the average energy of neutrinos and antineutrinos
  in $\gamma P$ hadronic interactions are presented in fig. \ref{fig4} versus 
  the photon energy as estimated from the neutrino yields.

\begin{figure}[!t]
 \centering
 \includegraphics[width=3.2in]{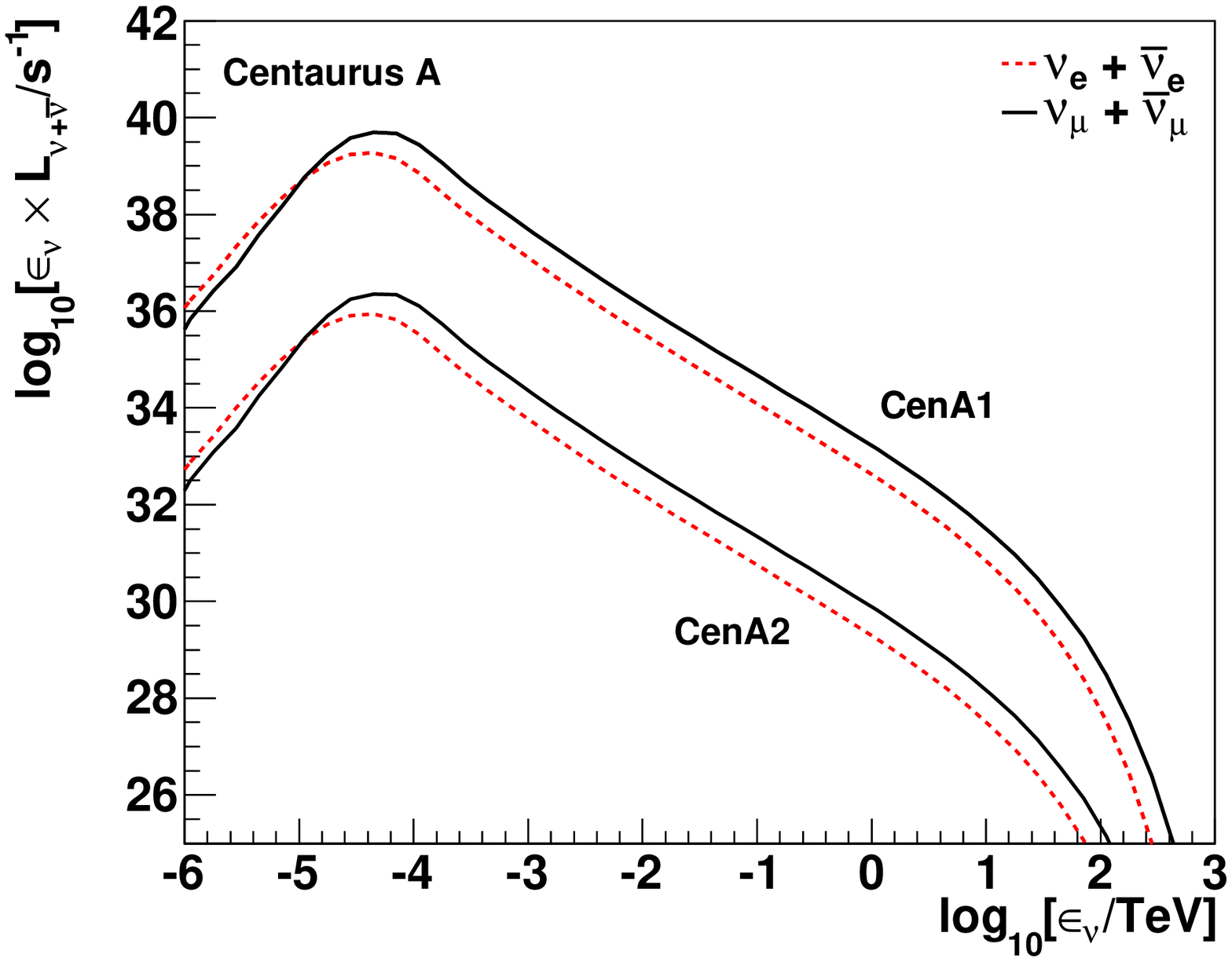}
 \includegraphics[width=3.2in]{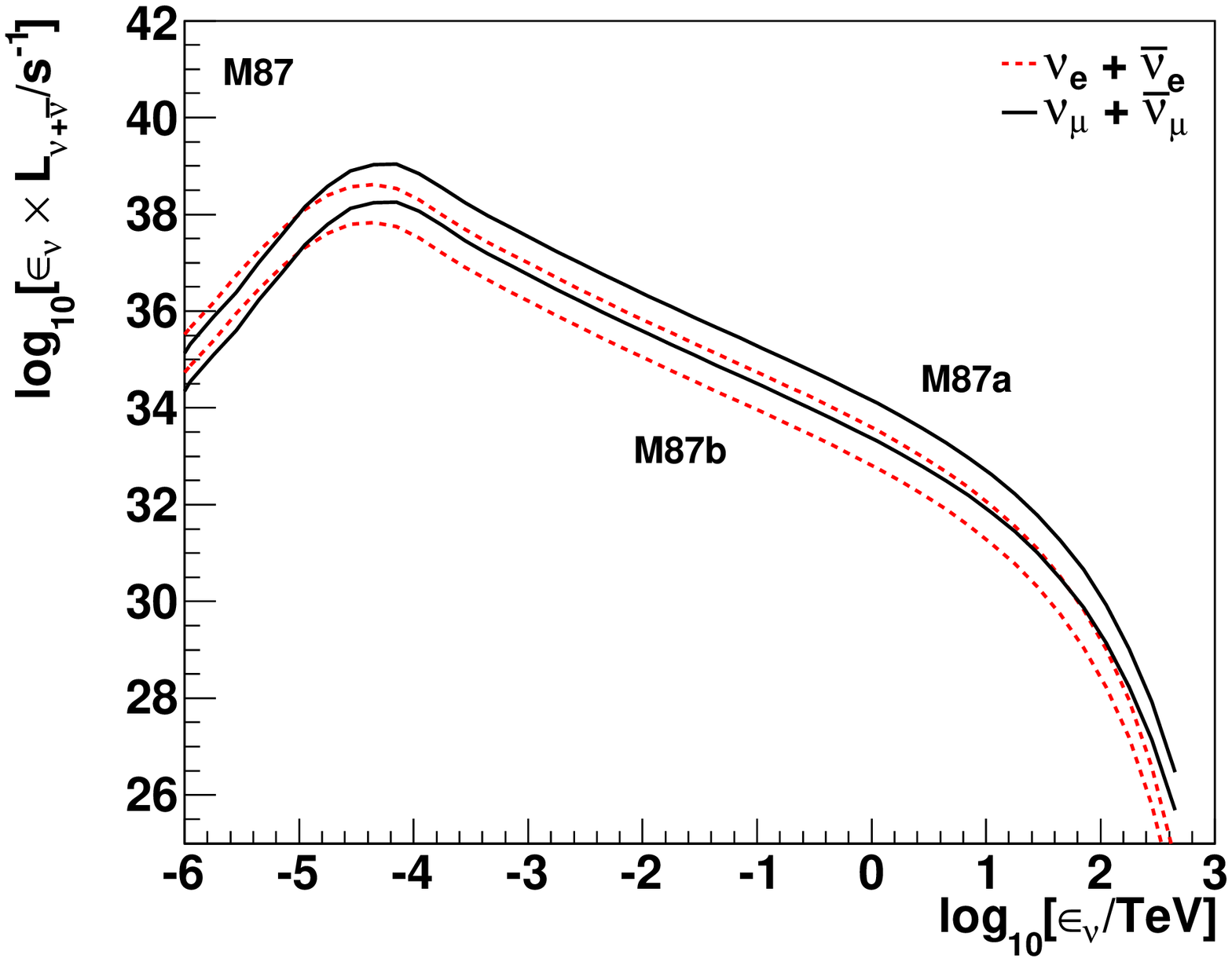}
  \caption{Electron (dotted line) and muon (solid line) neutrino luminosities
  expected from Centaurus A (left) and M87 (right) at source and produced by
  interactions of their own $\gamma$ radiation with the gas and dust in the
  corresponding galaxies. Different models for the gas and dust content in
  Centaurus A and M87 are considered (see text and table \ref{tab8}).}
 \label{fig5}
\end{figure}

 \section{Results}

 \subsection{Neutrino flux from CenA and M87}

 The electron and  muon neutrino luminosities at source for Centaurus A and M87, as 
 estimated with formula (\ref{eq2}), are presented in Fig. \ref{fig5}. Luminosities 
 are shown for different column depths of target protons: \textbf{CenA1},
 \textbf{CenA1}, \textbf{M87a} and \textbf{M87b}. It can be seen that neutrino
 luminosities exhibit a single power law behavior  only interrupted at low
 energies ($\epsilon_{\nu} \sim 10^{-4} \, \mbox{TeV}$) by a small bump that
 emerges as a result of the enhancement of the cross section due to the
 production of hadronic resonances in $\gamma P$ interactions. At
 high-energies, at approximately $\epsilon_\nu = 10 \, \mbox{TeV}$, the
 $L_{\nu + \bar{\nu}}$ functions are limited by an exponential cut, which is 
 generated by the equivalent one for the parent $\gamma$-ray luminosities 
 - see equation (\ref{eq9}). On the other hand, neutrino luminosities are
 restricted at low energies by the pion photoproduction energy threshold.

 The integrated neutrino luminosities above $100 \, \mbox{MeV}$ derived for 
 Centaurus A from fig. \ref{fig5} are $\mathcal{L}_{\nu + \bar{\nu}} (\epsilon_\nu >100 \, 
 \mbox{MeV}) \approx  10^{35} -  10^{32} \, \mbox{erg} \cdot \mbox{s}^{-1}$, for electron
 or muon neutrinos. Meanwhile, regarding the giant elliptical galaxy M87, it is found that 
 $\mathcal{L}_{\nu + \bar{\nu}} (\epsilon_\nu >100 \, 
 \mbox{MeV}) \approx  10^{35} -  10^{34} \, \mbox{erg} \cdot \mbox{s}^{-1}$.
 The above integrated luminosities are quite low values. 

\begin{figure}[!t]
 \centering
 \includegraphics[width=3.2in]{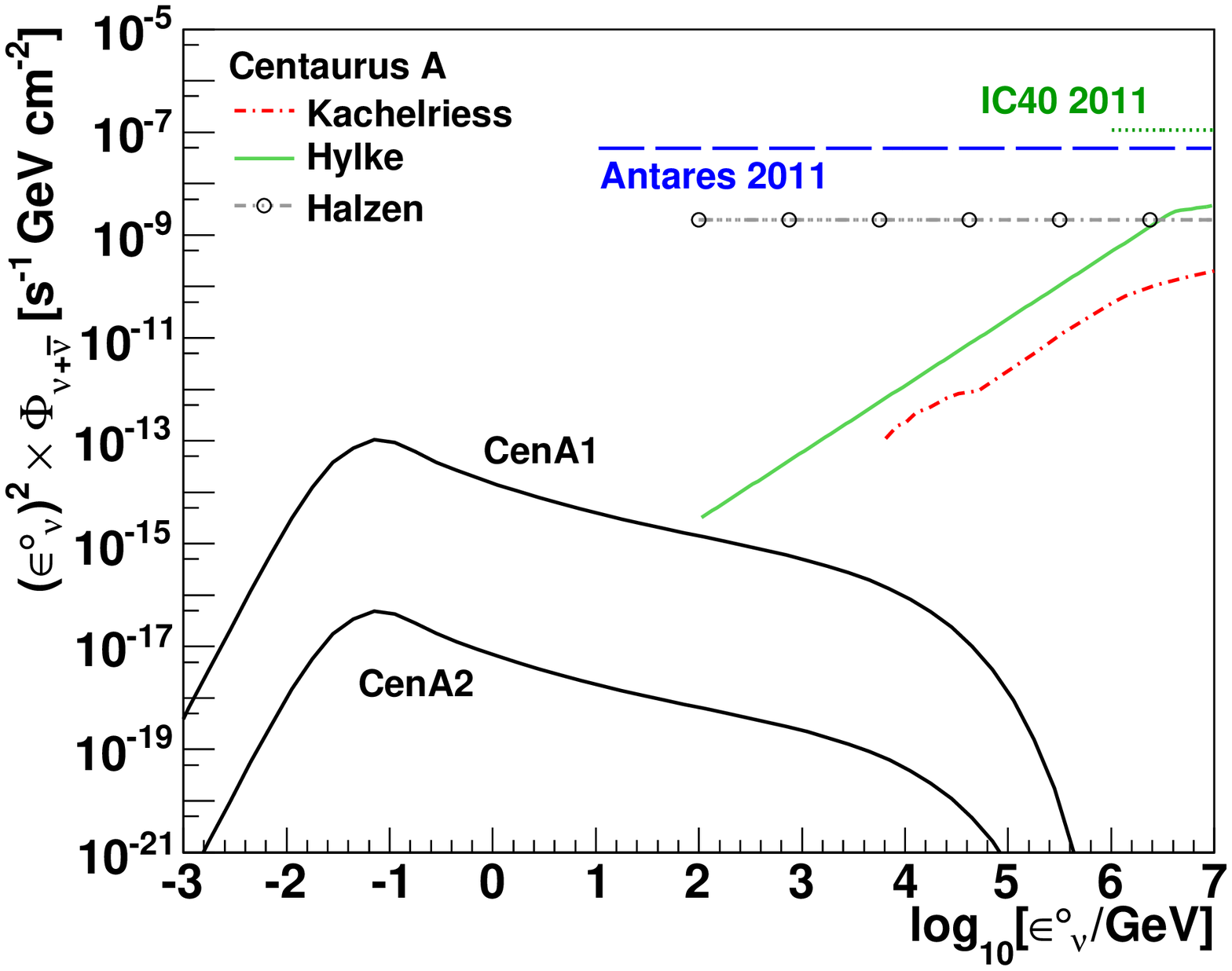}
 \includegraphics[width=3.2in]{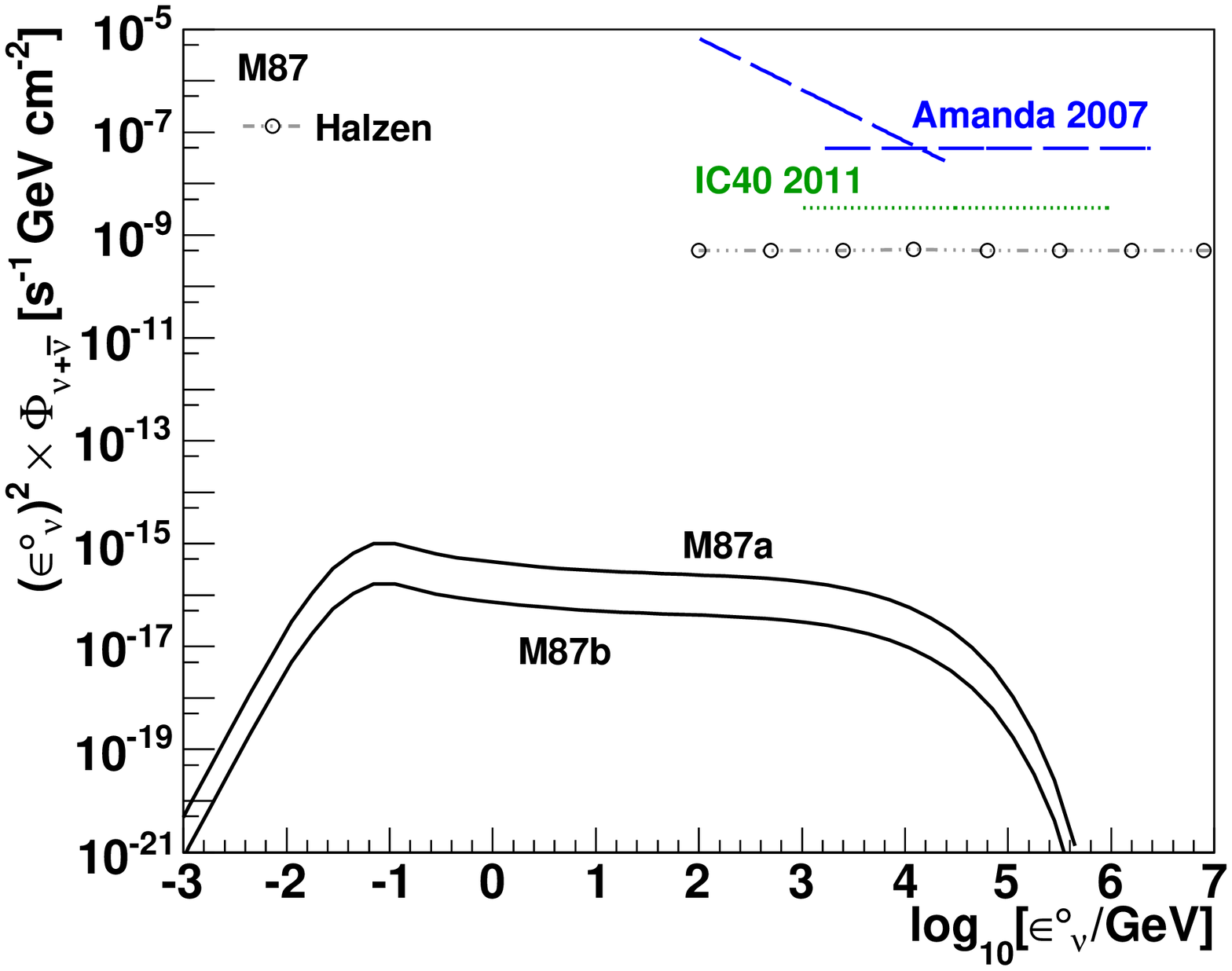}
  \caption{Neutrino and antineutrino fluxes expected from Centaurus A (left)
  and M87 (right) at Earth for each neutrino type. The flux is produced by
  interactions of their own $\gamma$ radiation with the gas and dust at the
  sources. Different models for the target material in Centaurus A and M87 are
  considered (see text). Neutrino oscillations are taken into
  account. Individual $90 \%$ C.L upper limits on the neutrino fluxes
  established with the Amanda \cite{amanda2007}, ICECUBE \cite{ic402011} and
  Antares \cite{antares2011} detectors for each galaxy are also
  shown. For M87, two different $\nu$ limits from Amanda are shown (segmented
  lines). They were estimated assuming different spectral indices
  for the differential $\nu$ flux: $\gamma = -2$ and $-3$ (horizontal and inclined
  lines, respectively) \cite{amanda2007}. Finally, predictions for Centaurus A
  based on three different hadronic models (Kachelriess \cite{Kachelriess09},
  Hylke \cite{Hylke08} and Halzen \cite{halzen08, halzen11}) are plotted
  (left) along with one theoretical estimation for M87 (right, Halzen
  \cite{halzen08, halzen11}).}
 \label{fig6}
\end{figure}

 From the particle luminosities at source, the flux of neutrinos observed at Earth
 can be estimated using the formula:
  \begin{equation}
  \Phi_{\nu + \bar{\nu}}(\epsilon^{\circ}_\nu) = \frac{(1+z)^2}{4\pi D^2_L} \cdot
  L_{\nu + \bar{\nu}}(\epsilon_\nu)
   \label{eq19}
 \end{equation}
 where $\epsilon^{\circ}_\nu = \epsilon_\nu/(1 + z)$ is the measured neutrino
 energy. If neutrino oscillations are taken into account \cite{nakamura2010}, 
 a flux ratio of the form $(\nu_e + \bar{\nu}_e)$:$(\nu_\mu +
 \bar{\nu}_\mu)$:$(\nu_\tau + \bar{\nu}_\tau)  = 1:1:1$ should be expected at
 Earth. Considering this effect upon propagation, the resulting neutrino
 fluxes, for each flavor, from Centaurus A and M87 are as shown in figure
 \ref{fig6}. In the same graph, the $90 \%$ confidence level upper bounds for
 neutrino fluxes from Centaurus A and M87 obtained by the
 Antares\cite{antares2011}, Amanda\cite{amanda2007} and ICECUBE\cite{ic402011}
 Collaborations are presented. From figure \ref{fig6}, it is clear that the
 predicted neutrino fluxes from $\gamma P$ interactions of high-energy
 radiation with gas and dust at the sources are too low to be observed, more
 than six orders of magnitude below the modern experimental limits for 
 the aforementioned AGN's.

 In figure \ref{fig6} (left panel), the expected $\nu$ fluxes from Centaurus A in 
 the framework of three different hadronic models are also shown for
 comparison. In all cases, the cosmic ray fluxes (composed by protons) are
 normalized using the data from the Pierre Auger observatory
 \cite{Auger07}. One model, due to Hylke et al., assumes that acceleration
 takes place at the jets \cite{Hylke08}. The second model, worked out by
 Kachelriess et al., invokes cosmic ray acceleration close to the core
 (with spectral index $\alpha = 2.7$) \cite{Kachelriess09}. The last model, 
 proposed in \cite{halzen08}, uses also the $\gamma$-ray data from
 the HESS detector \cite{Aharonian05} to put an upper limit on the neutrino
 flux assuming a pionic origin of the TeV radiation. This limit is also
 applied in \cite{halzen11} to M87 (see right panel of figure \ref{fig6}) by
 noting that luminosities from Centaurus A and M87 are similar at TeV
 energies. It is clear from Fig. \ref{fig6} that at high-energies, relevant
 for neutrino astronomy ($\epsilon_\nu \gtrsim 1 \, \mbox{TeV}$), neutrinos
 from the cosmic ray channel dominate over the modest contribution from
 $\gamma P$ interactions studied in this paper.

 \subsection{Diffuse neutrino flux from FRI galaxies}

\begin{figure}[!t]
 \centering
 \includegraphics[width=3.2in]{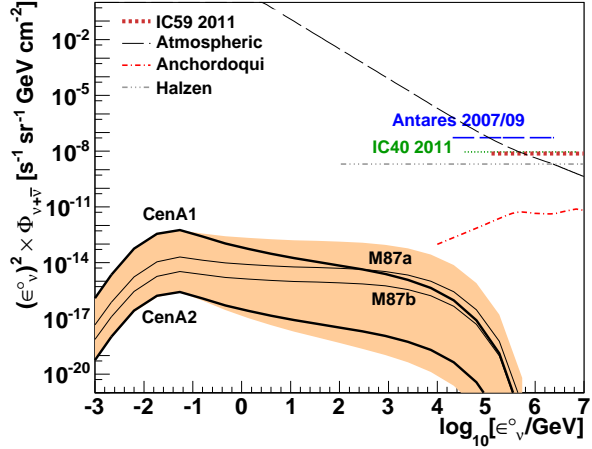}
  \caption{Diffuse flux of neutrinos (for each type) expected from FRI galaxies assuming
           different $\nu$ luminosities at the source. Individual luminosities
           are based on the predicted $L_{\nu + \bar{\nu}}$ functions of
           figure \ref{fig5} for Centaurus A (bold solid lines) and M87 (thin
           solid lines). Neutrino oscillation is taken into account. The $90
           \%$ C.L. upper limits on the diffuse neutrino flux derived by the
           ICECUBE \cite{ic402011b, ic592011} and the Antares
           \cite{antares2011b} collaborations are shown. For comparison, the
           diffuse flux of atmospheric neutrinos (segmented line)\cite{Gandhi98},
           including the prompt component from $D$-meson decays \cite{Volkova80},
           is also represented. Two predictions for the diffuse $\nu$ flux from AGN's 
           using hadronic  models (Anchordoqui \cite{Anchordoqui08} and Halzen 
           \cite{halzen08}) are included in the plot. The shadowed band covers 
           the results obtained by varying the  photon index (from $a = -2.08$
           to $-2.67$) and the angle-averaged column depths (FRI models of
           table \ref{tab8}).}
 \label{fig7}
\end{figure}

 The diffuse flux of neutrinos expected from FRI type galaxies, as derived
 from equation (\ref{eq4}), is shown finally in figure \ref{fig7}. Several
 curves are obtained, assuming that the individual $\nu$ luminosities of FRI
 type galaxies are similar to the ones estimated for Centaurus A or M87 (see
 figure 5). In this way, the diffuse neutrino fluxes here obtained share the
 same cuts and spectral shapes exhibited by the individual fluxes considered
 as reference models. The aforementioned results are compared in figure \ref{fig7} 
 with the corresponding $90 \%$ C.L. limits established with the Antares and 
 ICECUBE neutrino detectors. It is found that the predicted diffuse fluxes are
 more than seven orders of magnitude lower than the above experimental bounds
 inside the common energy range. Therefore, the estimated fluxes are 
 outside the reach of the modern neutrino telescopes.

 The diffuse flux derived for FRI type radio galaxies is also shown in figure 
 \ref{fig7} along with two predictions for the background flux of neutrinos
 from AGN's based on hadronic models with ultra high-energy protons
 \cite{halzen08} and iron nuclei \cite{Anchordoqui08}, respectively. It can be
 observed that the diffuse $\nu$ fluxes produced in the framework of hadronic
 models are dominating the high-energy range over the $\nu$ flux generated by
 interactions of the $\gamma$ radiation with gas and dust at source.

 \subsection{Diffuse neutrino flux from BL Lac galaxies}

\begin{figure}[!t]
   \centering
   \includegraphics[width=3.2in]{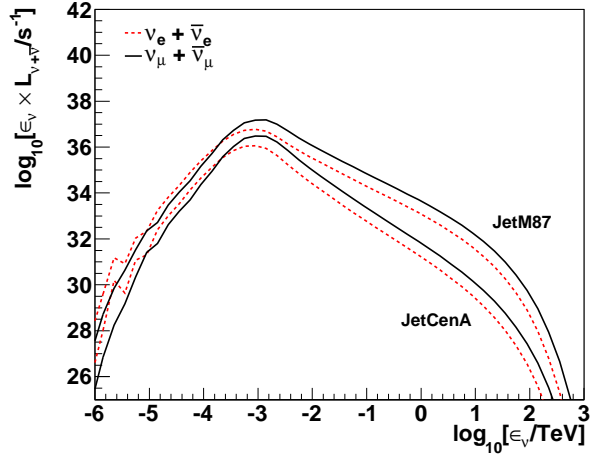}
   \caption{Electron (dotted line) and muon (solid line) neutrino luminosities
   at source expected from the BL Lac models \textbf{JetCenA} and
   \textbf{JetM87}.}
   \label{fig8}
 \end{figure}

 In figure \ref{fig8}, the $\nu_e$ and $\nu_\mu$ luminosities at source,
 calculated with the BL Lac models \textbf{JetCenA} and \textbf{JetM87}, are
 shown. They are estimated following relations (\ref{eq2}) and (\ref{eq3}). 
 As the FRI plots, the above graphs also display a depletion at high-energies
 (arising from the corresponding cut in the parent $\gamma$-ray spectrum), a
 bump and the effect of the pion photoproduction energy threshold at low
 energies. However, the last two features are slightly shifted to higher
 energies, when compared with the FRI luminosities, due to the fact that the
 target protons are moving outwards the gamma source at relativistic speeds
 ($\Gamma = 10$). In this regard, the bump is now located at $\epsilon_{\nu}
 \sim 10^{-3} \, \mbox{TeV}$.  Luminosities are lower in these BL Lac models
 than for their FRI counterparts. For the models \textbf{JetCenA} and
 \textbf{JetM87}, it is obtained that $\mathcal{L}_{\nu + \bar{\nu}}
 (\epsilon_\nu >100 \, \mbox{MeV}) \approx  10^{34} -  10^{33} \, \mbox{erg}
 \cdot \mbox{s}^{-1}$ and $\approx  10^{35} -  10^{34} \, \mbox{erg} \cdot
 \mbox{s}^{-1}$, respectively.

 The diffuse neutrino fluxes expected at Earth from BL Lac galaxies are shown in 
 figure \ref{fig9}. Estimations are performed for the models \textbf{JetCenA} 
 and \textbf{JetM87} and take into account neutrino oscillations. From fig. \ref{fig9},
 it can be noticed that the expected fluxes are more than eight orders of 
 magnitude below the present experimental limits at energies above $3 \times 10 \, 
 \mbox{TeV}$. That will very likely make these diffuse neutrino fluxes inaccessible for
 experiments like ICECUBE \cite{ic402011b, ic592011} or the future KM3NET \cite{Carr07}.

 \subsection{Uncertainties from the photon index}

\begin{figure}[!t]
 \centering
 \includegraphics[width=3.2in]{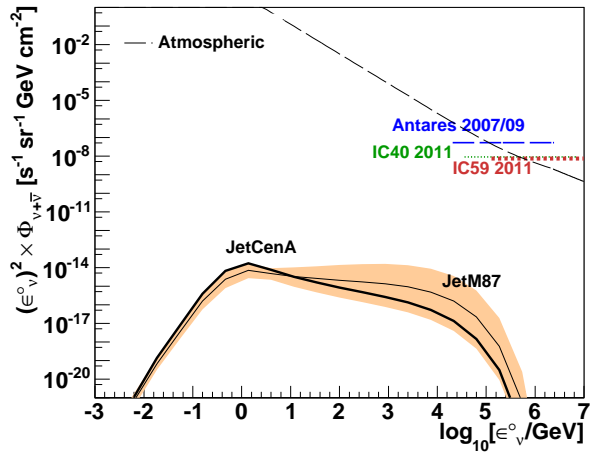}
  \caption{ Diffuse neutrino
  flux for each $\nu$ species from BL Lac type galaxies (models
  \textbf{JetCenA} and \textbf{JetM87}). The $90 \%$ C.L. upper limits on the
  diffuse neutrino flux derived by the ICECUBE \cite{ic402011b, ic592011} and the 
  Antares \cite{antares2011b} Collaborations are shown. For comparison, 
  the diffuse flux of atmospheric neutrinos (segmented line)\cite{Gandhi98},
  including the prompt component from $D$-meson decays \cite{Volkova80},
  is also presented. The shadowed band covers the results obtained by varying 
  the  photon index, $a$, from $-1.77$ to $-2.44$ and the angle-averaged
  column depths (BL Lac models of table \ref{tab8}).}
 \label{fig9}
\end{figure}

 One can still change the photon index, $a$, which appears in formula (\ref{eq9}), 
 inside the observable limits and modify the corresponding column depths  
 to obtain error bands for the fluxes from FRI and BL Lac galaxies. For example, for FRI 
 type objects
 $a = 2.39 \pm 0.28$ \cite{GLF-inoue11}. By varying this parameter inside the allowed
 interval (including the values reported in table \ref{tab1}) and by changing  
 the magnitudes of the angle-averaged column depth for FRI galaxies 
 (using the models \textbf{CenA1}, \textbf{CenA2}, \textbf{M87a} and \textbf{M87b}), 
 then the error band for the corresponding diffuse flux of FRI objects
 is obtained (see figure \ref{fig7}). In the most optimistic case, the diffuse flux 
 is of the order of ${(\epsilon^{\circ}_\nu)}^2 \Phi_{\nu + \bar{\nu}} \approx 10^{-13} \, 
 \mbox{s}^{-1} \, \mbox{sr}^{-1} \, \mbox{GeV} \, \mbox{cm}^{-2}$, however, 
 it is still too low to be detected. For the BL Lac population detected by 
 \textit{Fermi}-LAT, $a = 1.99  \pm 0.22 \, \mbox{(rm)}$ \cite{Fermi-list}.
 By repeating the above procedure for this kind of AGN's (in the framework of
 models \textbf{JetCenA} and \textbf{JetCenA2}), an error band is also 
 obtained for the respective diffuse neutrino flux (see figure
 \ref{fig9}). Considering this band, it is found that in the most optimistic
 case the diffuse flux for BL Lacs can reach values of the order of
 ${(\epsilon^{\circ}_\nu)}^2 \Phi_{\nu + \bar{\nu}} \approx 10^{-14} \,
 \mbox{s}^{-1} \, \mbox{sr}^{-1} \, \mbox{GeV} \, \mbox{cm}^{-2}$. As before,
 the situation is neither improved for this type of sources. 

 Therefore, the conclusion remains the same, the neutrino flux
 from $\gamma$-ray interactions in FRI and BL Lac galaxies is well out of the
 reach of current and future neutrino telescopes. But  
 even in the case that a large enough detector is ever built with sensitivity to 
 the limits quoted above, it will remain the problem of the atmospheric neutrino 
 background, which is orders of magnitude larger. To see a contribution from both 
 FRI and BL Lac neutrino fluxes against such a background will require a huge amount 
 of neutrino events. The same problem will be faced when looking for an excess of 
 gamma-ray induced neutrinos from the direction of the individual sources 
 Centaurus A and M87. That will require a very large amount of detected events.

 \section{Conclusions}

  The detection of gamma rays from several FRI and BL Lac galaxies together with
  the presence of gas and dust at the sources suggest that those objects should
  emit neutrinos through $\gamma P$ interactions and contribute to the
  neutrino background in the universe. However, it was shown that the
  corresponding diffuse neutrino flux is quite small (in the best case of the
  order of $E^2 \Phi_{\nu + \bar{\nu}} \lesssim 10^{-13} \mbox{s}^{-1} \,
  \mbox{sr}^{-1} \, \mbox{GeV} \, \mbox{cm}^{-2}$) and orders of magnitude
  below the atmospheric neutrino background. Therefore, it will
  escape observation from modern and future neutrino telescopes.   
  If the observed TeV radiation from FRI and BL Lac sources is of leptonic origin, 
  the diffuse neutrino flux presents a cut around $10 \, \mbox{TeV}$. 
  Additionally, this background is bounded at low energies by the pion photoproduction 
  energy threshold. Calculations were performed using as references the two nearest FRI 
  galaxies to the Earth: Centaurus A and M87, for the gamma ray spectra and the 
  gas and dust contents of FRI and BL Lac type objects.  The validity of the AGN 
  unification scenario was assumed. 

  Predictions for Centaurus A and M87 were also performed. Despite the proximity of 
  these objects, the present estimations indicate that the neutrino flux produced by 
  the channel here explored will be not observable at observatories like 
  ICECUBE or the KM3Net and in other future experiments. By modifying, in the present 
  calculations, the photon index of the individual sources inside the available 
  experimental ranges, the situation does not improve. In this way, if neutrinos are ever 
  detected from FRI and BL Lac galaxies, the $\nu$ production by the hadronic model is 
  favoured.

  Estimations are still dependent on the real position of the $\gamma$ ray source
  at the AGN and on the real shape of the $\gamma$-ray spectra in the inner
  regions of the galaxy. If energy losses of the parent leptonic or hadronic
  particles in the $\nu$-chain are negligible, the biggest contribution
  results when the source is located close to the massive object that feeds
  the AGN, hypothesis that at the moment it is not in contradiction with the
  multi-wavelength observations of the elliptical galaxy M87.

 \section*{Acknowledgments}

 The author thanks R. Engel for his suggestions on the Monte-Carlo programs for
 photo-hadronic interactions. J.C. Arteaga is also grateful to Claus Grupen
 and Arnulfo Zepeda for reading the manuscript and the helpful discussions to 
 improve it. The author also thanks the unknown reviewers for many useful 
 suggestions. This work has been partially supported by the
 Coordinaci\'on de la Investigaci\'on Cient\'\i fica de la Universidad
 Michoacana.

\end{document}